\documentclass[12pt, a4paper,fleqn]{article}
\usepackage[english]{babel}
\usepackage[T1]{fontenc}
\usepackage{nicefrac}
\usepackage{graphicx}
\usepackage[utf8]{inputenc}
\usepackage{wrapfig}
\usepackage{braket}
\usepackage{fancyhdr}
\usepackage{pifont}
\usepackage{amsmath}
\usepackage{amssymb}
\usepackage{amsfonts}
\usepackage{subcaption}
\usepackage{multicol}
\usepackage[toc,page]{appendix}
\usepackage{makecell}
\usepackage{geometry}
\usepackage{floatrow}
\usepackage{blindtext}
\usepackage{multicol}
\usepackage{comment}
\usepackage{listings}
\usepackage{color}
\usepackage{chngpage}
\usepackage{cancel}
\usepackage{mathrsfs}
\usepackage{dsfont}
\usepackage{tikz,pgfplots}
\usepackage{comment}
\usepackage{adjustbox}
\usepackage{pifont}

\usepackage{multirow}
\usepackage{url}
\usepackage{cite}

\pgfdeclarelayer{background}
\pgfsetlayers{background,main}

\pgfmathsetmacro{\myxlow}{-1}
\pgfmathsetmacro{\myxhigh}{1}
\pgfmathsetmacro{\myiterations}{6}

\usetikzlibrary{positioning}
\usetikzlibrary{arrows}

\usepackage{url}
\usepackage[hidelinks]{hyperref}

\hypersetup{
	colorlinks,
	citecolor=blue,
	filecolor=black,
	linkcolor=black,
	urlcolor=blue
}

\makeatletter
\newcommand\xleftrightarrow[2][]{%
	\ext@arrow 9999{\longleftrightarrowfill@}{#1}{#2}}
\newcommand\longleftrightarrowfill@{%
	\arrowfill@\leftarrow\relbar\rightarrow}
\makeatother

\newfloatcommand{capbtabbox}{table}[][\FBwidth]
\definecolor{dkgreen}{rgb}{0,0.6,0}
\definecolor{gray}{rgb}{0.5,0.5,0.5}
\definecolor{mauve}{rgb}{0.58,0,0.82}

\newcommand{\Q}{\mathcal{Q}}
\newcommand{\ZZ}{\mathbb{Z}}

\newcommand{\nn}{\nonumber}

\numberwithin{equation}{section}

\pgfplotsset{compat=1.18}
\begin{document}

	\begin{titlepage}
		\begin{center}
			\vspace{5mm}
			{\Huge{ \bf The landscape and the swampland of O(3,3) gauged supergravities \\}}
		\end{center}
		
		\begin{center}
			\vspace{5mm}
			{\textbf{ Mariana Gra$\tilde{\text{n}}\text{a}$, Diego Perugini}}
		\end{center}

			\begin{center}
			\vspace{3mm}
			{	 \textit{ Institut de Physique Théorique, Université Paris Saclay, CEA, CNRS Orme des
					Merisiers, 91191 Gif-sur-Yvette CEDEX, France}}
		\end{center}
		
        \vspace{10mm}
		\begin{abstract}
            \noindent
            Compactifications of string theory on group manifolds with fluxes, performed via the Scherk-Schwarz mechanism, give rise to gauged supergravities. Using double field theory, this procedure can be made T-duality covariant, yielding a set of geometric and non-geometric fluxes whose string theory origin is often unclear. In this paper, we systematically analyze the complete catalog of half-maximal gauged supergravities in seven dimensions. By constructing a dictionary between orbifold twists and generalized fluxes, we identify which gaugings admit a realization as (a)symmetric orbifolds, and which do not, establishing the latter as candidates for the gauged supergravity swampland.
                
        \end{abstract}

        \begin{flushleft}
			\vspace{130mm}
			{\small{Mariana.grana@ipht.fr, \,\,Diego.perugini@ipht.fr}}
		\end{flushleft}

	\end{titlepage}
	\tableofcontents

	\newpage
    
	\section{Introduction}
String compactifications generate a vast web of consistent lower-dimensional
theories, referred to as the string landscape. Determining the extent of this landscape is one of the central problems of string phenomenology. 
At low energies, a string compactification is captured by an effective field theory containing only the light degrees of
freedom, while the massive string and Kaluza-Klein states are integrated out or consistently
truncated. This effective description retains information such as the massless spectrum, gauge
symmetries and interactions, but generally loses part of the global and microscopic data of the
underlying string vacuum. In the swampland program (see \cite{brennan_string_2017,palti_swampland_2019,grana_herraez_2021,vanbeest_lectures_2021,agmon_lectures_2022} for reviews), one asks the converse question: given a low-energy
effective field theory, does it admit an ultraviolet completion in string theory, or more
generally in any consistent theory of quantum
gravity?
Ordinary consistency of the effective theory (e.g. anomaly cancellation) is
necessary but not sufficient to guarantee such a completion exists.

Gauged supergravities provide a natural framework for investigating whether a given gravitational
effective field theory admits a realization within string theory. Gauged supergravities are
deformations of ungauged supergravity theories in which a subgroup of the global symmetry group is
promoted to a local gauge symmetry. The gauging introduces minimal couplings, fermion mass terms, and
a scalar potential, while supersymmetry imposes strong consistency constraints on the allowed
deformations \cite{freedman_gauge_1977,freedman_n_1978,de_wit_n_1982,de_wit_n8_1982}. In the embedding-tensor formulation, the choice of gauge group and its
embedding into the duality group are encoded in a duality-covariant tensor subject to linear and
quadratic constraints~\cite{de_wit_lagrangians_2003,samtleben_lectures_2008,trigiante_gauged_2016}. These conditions select the representations compatible with
supersymmetry and ensure the closure of the gauge algebra. Since they can be formulated entirely
within the lower-dimensional theory, they make it possible to classify admissible gaugings without
specifying a compactification or any other ultraviolet origin. This procedure generates the full
space of consistent supergravity deformations.

 Some gaugings can be generated by compactifying string theory or higher-dimensional supergravity
in the presence of background fluxes, non-trivial internal geometry, or duality twists. Fluxes
threading cycles of the internal space give rise to lower-dimensional vacua characterized by
non-trivial charges, interactions, and scalar potentials (see \cite{grana_flux_2006,douglas_flux_2007,denef_physics_2007} for a
review). A paradigmatic example is provided by geometric Scherk--Schwarz reductions on group
manifolds: expanding the higher-dimensional fields in a left-invariant frame yields a consistent
truncation, with the structure constants of the underlying Lie algebra determining the gauging of
the lower-dimensional
theory \cite{scherk_spontaneous_1979,scherk_how_1979,hull_flux_2009,reid-edwards_flux_2009}.
These constructions illustrate how fluxes, twists and internal geometric data select particular
solutions of the embedding-tensor constraints. They do not, however, establish that every gauging
allowed by the lower-dimensional consistency conditions possesses a corresponding
compactification. Moreover, even among gaugings with a known higher-dimensional origin, the
distinction between flux and geometry is not invariant under string
dualities \cite{dabholkar_generalised_2006,hull_non-geometric_2009}.

Half-maximal supergravity in $D=10-d$ dimensions, coupled to $d$ vector multiplets, admits a
continuous $O(d,d)$ symmetry acting on both the fields and the gauging parameters. When the
theory arises from a toroidal compactification of string theory, the discrete subgroup
$O(d,d;\mathbb Z)$ is identified with the perturbative T-duality group, which exchanges, for
example, momentum and winding modes \cite{maharana_noncompact_1993,giveon_target_1994}. As a
result, compactification data such as background flux, internal geometry, or an outright T-duality twist, can all be organized into a
single duality-covariant framework. In particular, the components of the embedding tensor can be
interpreted in terms of the T-duality chain:
$$
    H_{abc}\longleftrightarrow
    \omega_{ab}^{\ \ c}\longleftrightarrow
    Q_{a}^{\ bc}\longleftrightarrow
    R^{abc},
$$
where $H$ denotes the NS--NS three-form flux, $\omega$ the geometric flux associated with a "twisted torus", and $Q$ and $R$ the so-called non-geometric fluxes \cite{Kachru:2002sk,Hellerman:2002ax,shelton_generalized_2007}.

While $H$-flux and geometric $\omega$ flux  admit conventional realizations in terms of flux compactifications and twisted or curved geometries, the string-theoretic interpretation of $Q$- and $R$-flux backgrounds is considerably less clear. Such fluxes are often inferred through putative T-duality transformations or identified through their contribution to the lower-dimensional gauging, but these descriptions do not provide a string construction \cite{wecht_lectures_2007,plauschinn_non-geometric_2019}. Determining which non-geometric flux configurations correspond to genuine string backgrounds therefore requires an intrinsically string-theoretic description.

(A)symmetric orbifolds provide exact world-sheet realizations of a broad class of geometric and non-geometric string backgrounds. They are constructed by quotienting a toroidal conformal field theory by a discrete symmetry acting on the left- and right-moving degrees of freedom of the string world-sheet \cite{dixon_strings_1985,dixon_strings_1986,narain_asymmetric_1987,narain_asymmetric_1991}. Since T-duality acts chirally on the world-sheet fields, transformations without a conventional geometric interpretation may nevertheless define exact symmetries of the conformal field theory at special points in moduli space. Combining a twist with a shift along an additional compact direction yields a freely acting (a)symmetric orbifold, which provides a string-theoretic realization of a compactification with a geometric or duality monodromy \cite{dabholkar_duality_2003,flournoy_nongeometry_2006}. 
The relation between orbifold constructions and gauged supergravities is, however, not automatic. Gauging parameters are defined over the continuous duality group and are constrained by supersymmetry and closure of the gauge algebra. By contrast, an orbifold twist must define a finite-order automorphism of the underlying CFT and admit a consistent action on the world-sheet Hilbert space. In addition, the twisted sectors must satisfy level matching and combine into a modular-invariant partition function \cite{dixon_strings_1986}. Orbifold realizability therefore imposes discrete and genuinely string-theoretic conditions that are not captured by the continuous classification of gauged supergravities.

Previous studies have established this correspondence for several explicit constructions. In particular, the gauge algebras of freely acting (a)symmetric orbifolds have been derived directly from their world-sheet description and related to backgrounds carrying geometric and non-geometric fluxes 
\cite{dabholkar_duality_2003,condeescu_asymmetric_2012,condeescu_gauged_2013}.  A closely
related programme has been carried out by Hull, Vandoren and collaborators in the context of
partial supersymmetry breaking, where quantization conditions on Scherk-Schwarz mass parameters
imposed by the string-theoretic orbifold uplift are used to identify which gauged supergravities
belong to the string landscape as opposed to the
swampland \cite{Hull:2020nsc,Gkountoumis:2023fol}.
Also the appearance of non-commutative and non-associative structure related to non-geometric fluxes has been studied \cite{Lust:2010iy,lust_t-duality_2010,Blumenhagen:2011ph,Blumenhagen:2011yv,condeescu_asymmetric_2012}. These examples demonstrate that orbifold CFTs can provide a microscopic string origin for gaugings involving $H$, $\omega$, $Q$ and $R$ fluxes. Nevertheless, a systematic characterization of which gauged supergravities admit such a realization is still missing. In particular, the inverse problem, starting from a complete classification of gaugings and determining which of them arise from consistent (a)symmetric orbifolds, has not been addressed exhaustively.

Half-maximal supergravity in seven dimensions provides a particularly suitable setting for addressing this problem. Coupled to three vector multiplets, the theory has continuous global symmetry $\mathbb{R}^+\times O(3,3)$, and its consistent deformations are encoded in a finite set of embedding-tensor components subject to quadratic constraints. The inequivalent duality orbits of these deformations have been classified in \cite{dibitetto_duality_2012,dibitetto_all_2015}. The resulting catalogue provides a controlled setting in which the subset admitting a realization as an (a)symmetric orbifold can be identified explicitly.

In this paper, we construct a map between type II (a)symmetric orbifold twists and the $O(3,3)$ generalized fluxes that encode the corresponding seven-dimensional gaugings. Applying this dictionary to the complete catalogue of \cite{dibitetto_all_2015}, we determine which duality orbits admit a symmetric-orbifold realization, which ones require a genuinely asymmetric construction, and those that cannot be obtained within this class of exact string backgrounds. Our analysis therefore provides a realization of the seven-dimensional orbifold landscape inside the larger space of consistent gauged supergravities. Gaugings lying outside its image constitute candidates for the gauged-supergravity swampland, although they may still admit string realizations beyond the orbifold framework considered here. Moreover, even within the duality
orbits that do admit an orbifold realization, only a handful of values of the continuous gauging
parameters entering the embedding tensor turns out to be reachable this way: string-theoretic
consistency, in the form of symmetries of the Narain lattice,  level matching and modular invariance, is so restrictive that it
singles out a finite set of values within each orbit, rather than merely a discrete infinity of
them, leaving the continuum of parameters in between in the swampland, at least within this class
of constructions.

The structure of the paper is as follows: in section 2 we review O(3,3) gauged supergravity theory and fix our conventions. We also quickly review the results of 
\cite{dibitetto_duality_2012,dibitetto_all_2015}, with particular emphasis on the gaugings and vacua classification.
In section 3 we review some basic aspects of Generalized Scherk-Schwartz reductions. 
While in section 4 we briefly review the basic aspects on non-geometric fluxes and discuss the already known string theory realizations of the gaugings in table \ref{tab:gaugings}.      
In section 5 we review orbifold constructions in a general fashion. 
We then specialize to $T^2\times S^1/\ZZ_N$ freely acting (a)symmetric orbifolds of type II superstring.
Computing the tree-level three point scattering amplitudes of the vectors in the orbifolded theory we build an explicit map between the orbifold twists and the $O(3,3)$ gaugings.
Section 6 contains the analysis of the fate of the section 2 vacua.
The section is complemented with several examples including the explicit partition functions of the various models. Modular invariance is tested for each construction explicitly, while integrality of the q-expation (i.e. the particle interpretation of the spectrum) is manifest due to the use of CFT characters.
In section 7 we summarize our results and make some comments on possible extensions of the present work.
This work is accompanied by various appendices, to which we defer some technical details.

	\section{\texorpdfstring{$O(3,3)$}{} gauged supergravity}

\label{sec:GSUGRA}

In this section we review the basics of O(3,3) gauged supergravity following \cite{dibitetto_duality_2012,dibitetto_all_2015} and set our notations. The ungauged theory can be obtained as a torus reduction of the ten-dimensional universal $\mathcal{N}=1$ gravitational sector of type I/II (or heterotic) supergravity. 
The 10d content of the theory is just the $\mathcal{N}=1$ sugra multiplet: a metric  tensor, an anti-symmetric two-form, a scalar, a left-handed gravitino and a right-handed dilatino :
\begin{equation}
	G_{10d}=\left\{	G_{_{MN}}, C_{_{MN}},\Phi	;\psi_{_{M,L}}	, \chi_{_R}\right\}.
\end{equation}
Its $T^3$ reduction yields, for what concerns the bosonic fields, a metric tensor, an anti-symmetric two-form, six vectors and ten scalars, while the fermionic degrees of freedom are one gravitino and four fermions. All these fields naturally organize themselves into a sugra multiplet and three vector multiplets in 7 dimensions:
\begin{equation}
	G_{7d}=\left\{g_{\mu\nu}, B_{\mu\nu},\Phi,3A_\mu;\psi_\mu,\chi\right\}, \qquad 3\times V_{7d}=3\times \left\{ A_{\mu},3\phi;\lambda \right\} \, .
\end{equation} 
Recall that in seven dimensions the minimal spinor is a symplectic Majorana spinor, which forms a doublet under the R-symmetry group and carries 8 on-shell degrees of freedom. All fermionic fields are of this type, but we have suppressed both spinor and symplectic indices to avoid cluttering of notation. The theory possesses 16 real off-shell supercharges,  $\mathcal{N}=1$ in 7d\footnote{This theory is sometimes called  $\mathcal{N}=2$ since all the spinorial fields are symplectic doublets.}, hence it is a half-maximal supergravity. 
The 7d theory posses a global symmetry group given by:
\begin{equation}
	G=\mathbb{R}^+_\Phi\times O(3,3)\equiv\mathbb{R}^+_\Phi\times G_0 \,.
\end{equation}
Here we will focus on gaugings that do not involve the ${\mathbb{R}^+}$ factor.  The six vectors of the theory are in the fundamental representation of $G_0$, 
while the 9 scalar fields parameterize the coset  $\frac{SO(3,3)}{SO(3)\times SO(3)}$. Using the isomorphism $SO(3,3)\simeq SL(4)$ one can explicitly parametrize the coset $\frac{SL(4)}{SO(4)}\simeq\frac{SO(3,3)}{SO(3)\times SO(3)}$ through the vielbein 
\begin{equation}
	E_m^{\ \bar{n}}=
	\begin{pmatrix}
			e^{\phi_1/2}&e^{\phi_2/2}\chi_1&e^{\phi_3/2}\chi_2&e^{-\left(\phi_1+\phi_2+\phi_3\right)/2}\chi_4\\
			0&e^{\phi_2/2}&e^{\phi_3/2}\chi_3&e^{-\left(\phi_1+\phi_2+\phi_3\right)/2}\chi_5\\
			0&0&e^{\phi_3/2}&e^{-\left(\phi_1+\phi_2+\phi_3\right)/2}\chi_6\\
			0&0&0&e^{-\left(\phi_1+\phi_2+\phi_3\right)/2}
	\end{pmatrix},
	\label{par}
\end{equation}
where $m$ is an $SL(4)$ index and $\bar{n}$ is an $SO(4)$ index (the set of conventions for the indices is given in \eqref{indices}). 
The above $SL(4)$ vielbein can be converted to an $O(3,3)$ generalized vielben via the t'Hooft symbols, and reads:
\begin{equation}
   \tilde{E}_A^{\   \tilde{B}}= \begin{pmatrix}
     e & b \, \hat e \\
     0 & \hat e
    \end{pmatrix} ,
\end{equation}
where $e$ and $\hat e$ are the veilbein of the metric and inverse metric respectively, and b is an antisymmetric field. The indices $\tilde{A},\tilde{B},\cdots$ are $SO(3)\times SO(3)$ indices.
The explicit form of the geometric vielbein on $T^3$ and the antisymmetric field can be given in terms of the field $\phi_i$ and $\chi_{_i}$ as:
\begin{equation}
    e=\begin{pmatrix}
    e^{(\phi_1+\phi_2)/2}&e^{(\phi_1+\phi_3)/2}\chi_3&e^{-(\phi_2+\phi_3)/2}\chi_5\\
        0&e^{(\phi_1+\phi_3)/2}&e^{-(\phi_2+\phi_3)/2}\chi_6\\
        0&0&e^{-(\phi_2+\phi_3)/2}
    \end{pmatrix},
\end{equation}
\begin{equation}
    b_{12}=\chi_4-\chi_1 \chi_5-\chi_2 \chi_6+\chi_1 \chi_3 \chi_6,
    \qquad
    b_{13}=-\chi_2+\chi_1 \chi_3,
    \qquad
    b_{23}=\chi_1.
\end{equation}
In this conventions the overall volume of $T^3$ is parameterized  by $e^{3\phi_1/2}$, while $\phi_2$ and $\phi_3$ parameterize ratios of radii.
The coset representative, or generalized metric, for the SL(4) parameterization  is given by: 
\begin{equation}
	M_{mn}=E_{m}^{\ \bar{m}}\ E_{n}^{\ \bar{n}}\ \delta_{\bar{m}\bar{n}} ,
\end{equation}
while for O(3,3) one has 
\begin{equation}
	\begin{split}
        \mathcal{M}_{AB}&=\frac{1}{2}M_{mn}\ M_{pq} \ \left[G_A\right]^{mp}\ \left[G_B\right]^{nq}
        \\
        &=\tilde{E}_{A}^{\ \tilde{C}}\ \tilde{E}_{B}^{\ \tilde{D}}\ \delta_{\tilde{C}\tilde{D}}. 
	\end{split}
\end{equation}
The kinetic terms for the scalar fields are:
\begin{equation}
	\begin{split}
	\mathcal{L}_{kin,scal.}=&-\frac{5}{2}\Phi^{-2}\left(\partial_\mu\Phi\right)^2+\frac{1}{2}\partial_\mu M_{mn}\partial^\mu M^{mn}\\
	&=-\frac{5}{2}\Phi^{-2}\left(\partial_\mu\Phi\right)^2+\frac{1}{4}\partial_\mu \mathcal{M}_{AB}\partial^\mu\mathcal{M}^{AB},
	\end{split}
	\label{kinscal}
\end{equation}
where $\mathcal{M}^{AB}$ is the inverse of $\mathcal{M}_{AB}$.
We now have all the ingredients to describe how to gauge a subgroup of $G_0$. 

\subsection{Gaugings}

The gauging procedure consists in promoting a subgroup $G'\subset G_0$ to a local symmetry gauged by the vectors of the theory.
The very first step is to introduce the so called \textit{embedding tensor} $\Theta$, which has an index in the fundamental and another one in the adjoint representation of $G_0$, and completely encodes all the information about the gauging: gauge group, new couplings, masses, etc. 
Concretely, $\Theta$ describes the embedding of $G'$ inside $G_0$, and in particular it builds the generators of $G'$ in terms of the generators of $G_0$. The closure of the gauge algebra requires a quadratic constraint  on $\Theta$.
Once the embedding tensor is introduced, one can substitute the derivatives of the scalar fields with covariant derivatives with respect to $G'$, 
namely 
\begin{equation} \label{covder}
    \partial_\mu \quad \to \quad D_\mu=\partial_\mu + g\,  A_\mu^{A}\,  X_{A}=\partial_\mu + g A_\mu^{[mn]} X_{[mn]},
\end{equation}
 where $g$ is the gauge coupling and the generators $X$ are built from the embedding tensor as follows (in SL(4) language) 
 \begin{equation}
     X_{[mn]}\equiv (\Theta_{mn})_{pq}{}^{rs}\ t^{pq}{}_{rs} \ ,
 \end{equation}
 where all pair of indices are antisymmetrized, and $t$ are the generators of the Lie algebra of $G_0$.    
Other modifications are needed to have a fully gauge-covariant theory, in particular one has to modify the field-strengths for the vectors and p-forms, and to introduce new topological couplings. This procedure will in general break supersymmetry, unless one introduces mass-like terms for the fermions and modifies the susy transformations. The consistency of this step requires an additional constraint on $\Theta$, the linear constraint. At this stage, the Lagrangian obtained is gauge and susy invariant up to first order in $\Theta$ (or equivalently in the gauge coupling $g$), while at second order in $g$ a scalar potential and new four fermions couplings appear. One can show that the quadratic constraint is sufficient to ensure that no higher order $g$ modifications are needed.
We will not review here all these steps, which can be found in e.g. \cite{samtleben_lectures_2008}, but only focus on the parts of the gauged Lagrangian that we  need.

Here we only consider gaugings entirely inside $O(3,3)$, and without an $O(3,3)$ singlet giving a St\"uckelberg-type coupling. For these gaugings, the linear constraint restricts the embedding tensor to the following SL(4) representations 
\begin{equation}
	\Theta \in 10\oplus 10' \implies \Q_{mn}:=\Q_{\left(mn\right)}, \quad \tilde{\Q}^{mn}:=\tilde{\Q}^{\left(mn\right)} \, .
	\label{LC}
\end{equation} 
These representations can be easily translated into $O(3,3)$ objects through the 't Hooft symbols in appendix \ref{App:conventions}, becoming the self-dual and anti self-dual parts of the $O(3,3)$ "flux" $f_{ABC}$:
\begin{equation} \label{fQ}
f_{ABC}^{^{+}}=\Q_{mn}\left[G_{A}\right]^{mp}\left[G_{B}\right]_{pq}\left[G_{C}\right]^{pn},\quad
	f_{ABC}^{^{-}}=\tilde{\Q}^{mn}\left[G_{A}\right]_{mp}\left[G_{B}\right]^{pq}\left[G_{C}\right]_{qn}.
\end{equation}
 The gauge algebra generators in \eqref{covder} are then given by:
\begin{equation} \label{Xmnpqrs}
	\left(X_{mn}\right)_{pq}^{\ \ rs}=\frac{1}{2}\left(\delta^{\left[r\right.} _{\left[m\right.}	\Q_{\left.n\right]\left[p\right.}	\delta^{\left.s\right]}_{\left.q\right]}+\frac{1}{2}\epsilon_{tmn\left[p\right.}\tilde{\Q}^{t\left[r\right.}\delta_{\left.q\right]}^{\left.s\right]}
	\right),
\end{equation}
or in the $O(3,3)$ language as:
\begin{equation}
	f_{AB}^{\ \ \ \; C}=f_{ABD}\ \eta^{DC}=	\left[\left(X_{mn}\right)_{pq}^{\ \ rs} \ \left[G_A\right]^{nm}	 \ \left[G_B\right]^{qp}  \ \left[G_D\right]_{rs}
	\right]\ \eta^{DC} \, ,
\end{equation}
where the $O(3,3)$ invariant metric is  $\eta=\begin{pmatrix}
    \mathbf{0}_{3}&\mathds{1}_3\\
    \mathds{1}_3&\mathbf{0}_{3}
\end{pmatrix}$, and it rises and lowers the indices $A,B,C,\cdots$ indices. 
The quadratic constraint ensuring the closure of the gauge algebra\footnote{As for the linear constraint this is true for gaugings only in the $O(3,3)$ part of $G$.}  reads:\begin{equation}
	\tilde{\Q}^{ml}\ \Q_{ln}=\frac{1}{4}	 \ \tilde{\Q}^{ls}\ \Q_{sl}\ \delta^m_n,
	\label{QC}
\end{equation}
which implies that the product of $\Q$ and $\tilde{\Q}$ is pure trace. 
This can be rewritten in terms of $f_{ABC}$ as follows:
\begin{equation} \label{Jacobi}
	f_{D\left[A\right.}^{\ \ \ \ E} f_{\left.BC\right]}^{\ \ \ \ D}=0.
\end{equation}
By means of duality transformations, $\Q$ and $\tilde \Q$ can be brought  at the same time to a diagonal form
\begin{equation}
\Q={\rm diag}(\alpha_1,\alpha_2,\alpha_3,\alpha_4) \, , \quad \tilde \Q={\rm diag}(\beta_1,\beta_2,\beta_3,\beta_4) \ .
\label{eq:QQt}
\end{equation}
The commutation relations for the generators $X_{mn}$, dictated by \eqref{Xmnpqrs} then simplify to 
\begin{equation} 
    [X_{mn},X_{pq}]=  
    \left( \frac{1}{2}\alpha_l\,\delta^r_{\left[m\right.}\delta^{l}_{n}\delta_{\left.n\right]\left[\right.p}\delta^s_{\left.q\right]}
    +\frac{1}{4}\beta_r\,\epsilon^r_{\ mn\left[p\right.}\delta^s_{\left.q\right]}
    \right)X_{rs} \, .
    \label{commXmn}
\end{equation}
In $O(3,3)$ language, these commutation relations read simply: 
\begin{equation}
    \left[X_A,X_B\right]=f_{AB}^{\ \ \ C} \ X_C.
    \label{eq:so33gen}
\end{equation}
Decomposing the $O(3,3)$ generators into  $SO(3)\times SO(3)$ components as $X_A=(Z_a,\chi^a)$, with $a=1,2,3$ fundamental $SO(3)$ indices, and similarly for the fluxes, the algebra reads
\begin{equation}
	\begin{split} \label{commutators}
		\left[Z_a,Z_b\right]&=H_{abc}\;\chi^c+\omega_{ab}^{\ \ \ c}\;Z_c,
		\\
		\left[Z_a,\chi^b\right]&=-\omega_{ac}^{\ \ \ b}\;\chi^c+Q_{a}^{\ \ bc}\;Z_c,
		\\
		\left[\chi^a,\chi^b\right]&=Q_{c}^{\ ab}\;\chi^c+R^{abc}\;Z_c \, .
	\end{split}
\end{equation}
These fluxes are related to the components of $\Q$ and $\tilde{Q}$ in \eqref{eq:QQt} by
\begin{equation}
	\Q=\text{diag} \left(H_{123},Q_{1}^{\ 23},Q_{2}^{\ 31},Q_{3}^{\ 12}\right),
	\quad
	\tilde{\Q}=\text{diag} \left(R^{123},\omega_{23}^{\ \ 1},\omega_{31}^{\ \ 2},\omega_{12}^{\ \ 3}\right).
    \label{Q-HwQR}
\end{equation}

Fluxes generate a scalar potential, and mass terms for the fermions. Up to an irrelevant overall coefficient, the latter are:
\begin{equation}
	 {\cal L_{\rm f. \, mass}}\ \sim g\left(A_1^{ij}\, \bar{\psi}_{\mu,i}\, \gamma^{\mu\nu}\,\psi_{\nu,j}+A_2^{ij}\,\bar{\psi}_{\mu,i}\,\gamma^\mu\,\chi_j+A_{3,\hat{\imath}\hat{\jmath}k}^{\quad \, \ l}\,\bar{\psi}_{\mu,l}\,\gamma^\mu\, \lambda^{k (\hat{\imath}\hat{\jmath})}+h.c.\right)
\end{equation}
where we have explicitly written  the fermionic symplectic $SU(2)_R$ fundamental representation indices $i,j,...$ (see Appendix \ref{App:conventions}), and also used the fact that the three $\lambda$ fermions can be thought as fermions in the symmetric representation\footnote{This choice is clear if one thinks of $SO(4)$ in the coset as $SU(2)_R\times SU(2).$} of $SU(2)$ labelled by $(\hat{\imath}\hat{\jmath})$, with $\hat{\imath},\hat{\jmath},\dots$ indices of the fundamental of $SU(2)$. 
The explicit form of the "shift matrices" $A_{1,2,3}$ is better expressible in terms of the vielbein with $SU(2)_R\times SU(2)$ indices:
\begin{equation}
	E_{m}^{\ i\hat{\imath}}=\frac{1}{\sqrt{2}} \ E_{m}^{\ \bar{n}} \left[\Gamma_{\bar{n}}\right]^{i\hat{\imath}}, \qquad E^{m}_{\ i\hat{\imath}}=\frac{1}{\sqrt{2}} \ E^{m}_{\ \bar{n}} \left[\bar{\Gamma}^{\bar{n}}\right]_{i \hat{\imath}} \ ,
\end{equation}
with $\Gamma$ the Dirac matrices of SO(4).
One finds the following expressions:
\begin{equation}
	\begin{split}
	A_1^{ij}&=\frac{1}{8}\Phi^{-1}\left(\Q_{mn}\ E^{m}_{\  k \hat{\imath}}E^{n}_{\ l \hat{\jmath}}\ \epsilon^{k i}\epsilon^{l j}\epsilon^{\hat{\imath} \hat{\jmath}}+\tilde{\Q}^{mn}\ E_{m}^{\ i \hat{\imath}}E_{n}^{\ j \hat{\jmath}}\epsilon_{\hat{\imath}\hat{\jmath}}\right),
	\\
	A_2^{ij}&=-\frac{1}{4}A_1^{ij},
	\\
	A_{3,\hat{\imath}\hat{\jmath}k}^{\quad \, \ l}&=\frac{1}{8} \Phi^{-1}\left(\Q_{mn}\ E^m_{\ k \left( \hat{\imath} \right.}\	E^n_{\ i \left. \hat{\jmath}\right)}	\ \epsilon^{i l}	-\tilde{\Q}^{mn}\ E_m^{\ i k}E_{n}^{\ l j} \epsilon_{i k}\ \epsilon_{k\left(\hat{\imath}\right.} \epsilon_{\left. \hat{\jmath}\right) j}			\right).
	\end{split}
\end{equation}
The scalar potential can be written as follows:
\begin{equation}
\label{VSL4}
	\begin{split}
	V=&\frac{g}{64}\Phi^{-2}\left(\frac{1}{4}\Q_{mn}\Q_{pq}\left(2M^{mp}M^{nq}-M^{mn}M^{pq}\right)+\tilde{\Q}^{mn}\Q_{mn}\right.
	\\
	&\qquad \quad \ \ 	\left.+\frac{1}{4}\tilde{\Q}^{mn}\tilde{\Q}^{pq}\left(2M_{mp}M_{nq}-M_{mn}M_{pq}\right)\right),
	\end{split}
\end{equation}
or equivalently as:
\begin{equation}
		V=\frac{g}{64}\Phi^{-2}\left(\frac{1}{6}f_{AC}^{\ \ \ E}f_{BD}^{\ \ \ F}\mathcal{M}^{AB}\mathcal{M}^{CD}\mathcal{M}_{EF}+\frac{1}{2}f_{DA}^{\ \ \ C}f_{CB}^{\ \ \ D}\mathcal{M}^{AB}+\frac{1}{3}f_{ABC}f^{ABC}\right).
		\label{gauged_pot}
\end{equation}
Recalling that the $SU(2)_R$ ($SU(2)$) indices are raised and lowered by $\epsilon_{ij}$ ($\epsilon_{\hat{\imath}\hat{\jmath}}$), the above potential can be rewritten in terms of the fermion shift matrices as follows:
\begin{equation}
	\begin{split}
			V&=g^2\left(-\frac{3}{10}|A_1|^2+\frac{4}{5}|A_2|^2-\frac{1}{2}|A_2|^2\right)\\
			&=-\frac{g^2}{4}\left(|A_1|^2+2|A_3|^2\right) \ ,
	\end{split}
\end{equation}
where  the first equality holds for any kind of gauging, whereas the second holds only for gaugings purely in $O(3,3)$. 

The last information we need are the supersymmetry conditions for the vacua, which require the equality \cite{samtleben_lectures_2008}:
\begin{equation}
	A_1^{ij}\ \chi_{j}=\sqrt{-\frac{10}{3}\ \frac{V}{16 \ g}} \ \chi^{i},
\end{equation}
to be satisfied for an arbitrary $SU(2)_R$ doublet spinor $\chi_i$. Notice that the 16 in the denominator is the real dimension of the spinors $\chi_i$, that is the number of supercharges. 

The list of inequivalent solutions to the quadratic constraint \eqref{QC}, found in \cite{dibitetto_duality_2012} \footnote{Two gauging are said equivalent if their are related by a T-duality transformation $O(3,3;\mathbb{Z})$.},
are eleven families depending on two parameters $\alpha$ and $\beta$, and are reported in Table \ref{tab:gaugings}. 
\begin{table}
\begin{adjustbox}{width=\textwidth}
	\begin{tabular}{| c | c | c | c |}
				\hline
				\textrm{ID} & $(H_{123},Q_{1}^{\ 23},Q_{2}^{\ 31},Q_{3}^{\ 12})$ & $(R^{123},\omega_{23}^{\ \ 1},\omega_{31}^{\ \ 2},\omega_{12}^{\ \ 3})$ &  gauging \\[1mm]
				\hline \hline
				$1$ & $\alpha$ diag($1,1,1,1$) & $\beta$ diag($1,1,1,1$) &  $\begin{cases}\textrm{SO}(4) , & \alpha\neq \pm \beta,\\ \textrm{SO}(3) \times U(1)^3 , & \alpha = \pm\beta.\end{cases}$\\[4mm]
				\hline
				$2$ & $\alpha$  diag($1,1,1,-1$) & $\beta$ diag($1,1,1,-1$) &  SO($3,1$)\\[1mm]
				\hline
				$3$ &$\alpha$  diag($1,1,-1,-1$) & $\beta$ diag($1,1,-1,-1$) &  $\begin{cases}\textrm{SO}(2,2) , & \alpha \neq \pm 
                \beta,\\ \textrm{SO}(2,1) \times U(1)^3 , & \alpha = \pm\beta.\end{cases}$\\[2mm]
				\hline
				$4$ & $\alpha$ diag($1,1,1,0$) & $\beta$ diag($0,0,0,1$)  &
                $\begin{cases}
                \textrm{ISO}($3$) , & \alpha \neq 0, \\
                \textrm{CSO}(1,0,3) , & \alpha=0,\beta\neq0.
                \end{cases}$\\[4mm]
				\hline
				$5$ & $\alpha$ diag($1,1,-1,0$) & $\beta$ diag($0,0,0,1$) & 
                $\begin{cases}
                \textrm{ISO}(2,1) , & \alpha \neq 0, \\
                \textrm{CSO}(1,0,3) , & \alpha=0,\beta\neq0.
                \end{cases}$\\[4mm]
				\hline
				$6$ & $\alpha$ diag($1,1,0,0$) & $\beta$ diag($0,0,1,1$)  & $\begin{cases}\mathfrak{s}^{ \frac{|\alpha|-|\beta|}{|\alpha|+|\beta|}}_{6,3^+} , & \alpha \neq \pm \beta, \\
                \mathfrak{s}_{6,2^+} , & \alpha = \pm\beta,
                \\
                \textrm{CSO}(2,0,2) , & \beta=0 \vee \alpha =0 .\end{cases}$\\[4mm]
				\hline
				$7$ & $\alpha$ diag($1,1,0,0$) & $\beta$ diag($0,0,1,-1$) & $\begin{cases}
                
                \mathfrak{s}_{6,4}   , & \alpha,\beta\neq0,
                \\
                \textrm{CSO}(1,1,2), & \alpha =0,
                \\
                \textrm{CSO}(2,0,2) , & \beta=0 .\end{cases}$\\[4mm]
				\hline
				$8$ & $\alpha$ diag($1,1,0,0$) & $\beta$ diag($0,0,0,1$) &
                $\begin{cases}
                \mathfrak{s}_{6,1^+} , & \alpha,\beta \neq 0, \\
                \textrm{CSO}(1,0,3) , & \alpha=0 ,\\ \textrm{CSO}(2,0,2) , &\beta =0 .\end{cases}$\\[4mm]
				\hline
                $9$ & $\alpha$ diag($1,-1,0,0$) & $\beta$ diag($0,0,1,-1$) & $\begin{cases}
                \mathfrak{s}^{ \frac{|\alpha|-|\beta|}{|\alpha|+|\beta|}}_{6,3^-} , & \alpha \neq \pm \beta, \\
                \mathfrak{s}_{6,2^-}, & \alpha = \pm\beta,
                \\
                \textrm{CSO}(1,1,2) , & \beta=0 \vee \alpha =0 .\end{cases}$\\[4mm]
				\hline
				$10$ & $\alpha$ diag($1,-1,0,0$) & $\beta$ diag($0,0,0,1$) &$\begin{cases}
                \mathfrak{s}_{6,1^-} , & \alpha,\beta \neq 0, \\
                \textrm{CSO}(1,0,3) , & \alpha=0 ,\\ \textrm{CSO}(1,1,2)  , &\beta =0 .\end{cases}$\\[4mm]
				\hline
                $11$ & $\alpha$ diag($1,0,0,0$) & $\beta$ diag($0,0,0,1$) &
				$\begin{cases} \mathfrak{n}_6 , &\alpha,\beta \neq 0  \ ,\\
                \textrm{CSO}(1,0,3) , &\alpha \vee \beta =0. 
                \end{cases}$\\[4mm]
				\hline
			\end{tabular}
    \end{adjustbox}
    \caption{
    All the T-duality orbits of consistent gaugings in half-maximal supergravity in $D=7$ \cite{dibitetto_duality_2012,dibitetto_all_2015}. The notation $\mathfrak{s}^{\gamma}_{6,i}$ indicates a one-parameter family of six-dimensional solvable algebras, that we label with the index $i$, while $\mathfrak{n}_6$ is a nilpotent six-dimensional algebra. The explicit form of the algebras is given in Appendix \ref{app:algebras}. }
    \label{tab:gaugings}
\end{table}

\subsection{Vacua}

At this point we have all the ingredients to study the vacua.
Because the moduli space is homogeneous, one can bring without loss of generality any vacua to a reference point, "the origin", as developed in \cite{dibitetto_charting_2011}. This corresponds to all fields $\phi,\chi$ set to 0, such that the vielbein \eqref{par} becomes the identity matrix. We list the vacua\footnote{Because of the contribution of the dilaton to the potential by an overall factor, see \eqref{VSL4}, the models where $V$ at the origin is not zero are not actually vacua.} in Table \ref{tab:spectra}, together with the (diagonalized) boson mass matrix. We labeled the scalar spectrum, inside squared brackets, by the mass of the states and with the multiplicity  of that state as a subscript.
We show in the next subsection the details of the procedure to search for vacua  by means of an example. Notice that in some cases (e.g. 7,8) only a specific configuration of fluxes leads to a vacuum (e.g. $\beta=0$).
\begin{table}
\begin{adjustbox}{width=\textwidth}
			\renewcommand{\arraystretch}{1.6}
			\begin{tabular}{| c | c | c | c | c | c |}
				\hline
				ID& $(H_{123},Q_{1}^{\ 23},Q_{2}^{\ 31},Q_{3}^{\ 12})$ & $(R^{123},\omega_{23}^{\ \ 1},\omega_{31}^{\ \ 2},\omega_{12}^{\ \ 3})$& V$|_{Origin}$& Scalars $M^2$&Susy \\
				\hline \hline
				1&$\alpha$ diag$\left(1,1,1,1\right)$& $\beta$ diag$\left(1,1,1,1\right)$&$-2\left(\alpha-\beta\right)^2$&$\left[\left(0\right)_{9}\right]$&\checkmark$\left(\alpha=\beta\right)$\\
				\hline
				2&$\alpha$ diag$\left(1,1,1,-1\right)$& $\pm\alpha$ diag$\left(1,1,1,-1\right)$&$2(1\pm2)\alpha^2$&$\left[\left(2\alpha^2\right)_{6},(-)_3\right]$&$\times$\\
				\hline
				3&$\alpha$ diag$\left(1,1,-1,-1\right)$& $\beta$ diag$\left(1,1,-1,-1\right)$&2$\left(\alpha+\beta\right)^2$&$\left[0,\left(2\left(\alpha^2+\beta^2\right)\right)_{4},(-)_4\right]\footnote{For $\alpha=\beta$ the eaten bosons are two instead of four.}$&\checkmark$\left(\alpha=-\beta\right)$\\
				\hline
				4&$\alpha$ diag$\left(1,1,1,0\right)$& $\pm\alpha$ diag$\left(0,0,0,1\right)$&$-\frac{\alpha^2}{2}$&$\left[\left(\alpha^2/2\right)_{6},(-)_3\right]$&$\times$\\
				\hline
				5&$\alpha$ diag$\left(1,1,-1,0\right)$& $\beta$ diag$\left(0,0,0,1\right)$&-&-&No Vacua\\
				\hline
				6&$\alpha$ diag$\left(1,1,0,0\right)$& $\beta$ diag$\left(0,0,1,1\right)$&0&$\left[0,\left(\alpha^2\right)_{2},\left(\beta^2\right)_{2},(-)_4\right]$\footnote{For $\alpha=\pm\beta$ the eaten bosons are two instead of four.}&$\checkmark$ $\left(\alpha=\beta\right)$ \\
				\hline
				7&$\alpha$ diag$\left(1,1,0,0\right)$&  $0\times$diag$\left(0,0,1,-1\right)$&0&$\left[(0)_3,\left(\alpha^2\right)_{2},(-)_4\right]$&$\times$\\
				\hline
				8&$\alpha$ diag$\left(1,1,0,0\right)$&  $0\times$diag$\left(0,0,0,1\right)$&0&$\left[(0)_3,\left(\alpha^2\right)_{2},(-)_4\right]$&$\times$\\
				\hline
				9&$\alpha$ diag$\left(1,-1,0,0\right)$& $\beta$ diag$\left(0,0,1,-1\right)$&-&-&No Vacua\\
				\hline
				10&$\alpha$ diag$\left(1,-1,0,0\right)$& $\beta$ diag$\left(0,0,0,1\right)$&-&-&No Vacua\\
				\hline
				11&$\alpha$ diag$\left(1,0,0,0\right)$& $\beta$ diag$\left(0,0,0,1\right)$&-&-&No Vacua\\
				\hline
			\end{tabular}    
    \end{adjustbox}
	\caption{Scalar spectra for each model, and values of the potentials. A dash  - means that the corresponding number of scalars are eaten by massless vector bosons to get mass. Gaugings 7 and 8 have a vacuum only for $\beta=0$. } 
	\label{tab:spectra}
\end{table}
Let us point out that without gauging the ${\mathbb R}^+$ factor there is no way to stabilize the dilaton. The vacua for which $V=0$ at the origin are therefore of no-scale type, and are the ones which could possibly be uplifted to string theory without further ingredients.
Other extrema of the potential at the origin (such as e.g. gauging 1 with $\alpha \neq \beta$) would then require stabilising the dilaton either beyond the supergravity regime, or breaking Lorentz invariance in the external space\footnote{As we will see, some of these gaugings come from reductions on $S^3$ with tree-form flux on the sphere. These can be uplifted to string theory either by coupling to a linear dilaton, or by adding three-form flux on the external space such that the vacuum becomes $(AdS_3 \times T^4) \times S^3$.}. 
We work out the precise string theory realization of the Minkowski vacua in section \ref{sec:landscape_swampland}.

\subsection{An explicit example}
\label{sec:exemplevacua}
To be more concrete, we show in detail the vacuum corresponding to the entry number 6 in Table \ref{tab:spectra}.  The massless scalars, vectors and fermions depend on the specific values of $\alpha,\beta$.
Focusing on the scalars, one can write down their kinetic term, considering the parametrization \eqref{par}, as:
\begin{equation}
	\mathcal{L}_{kin,scalars} \supset  -\frac{1}{2}K_{ij}\ \partial_\mu \varphi^i \ \partial^\mu \varphi^j,
\end{equation}
where $\varphi=\left(\phi_1,\phi_2,\phi_3,\chi_1,\chi_2,\chi_3,\chi_4,\chi_5,\chi_6\right)$, and we ignore the dilaton.
We have:
\begin{equation}
    \left.K_{ij}\right|_{\textrm{$\varphi$*}} \ =\left(
\begin{array}{c|c}
	\begin{array}{ccc}
		2 & 1& 1 \\
		1 & 2 & 1  \\
		1 & 1 & 2  \\
	\end{array}
	&
	\mathbf{0}_{_{3\text{x}6}}\\
	\hline
	\mathbf{0}_{_{3\text{x}6}}& 2\, \mathds{1}_{_{6\text{x}6}}\\
\end{array}
\right)
\label{k}
\end{equation}
\noindent
at the origin of moduli space, $\varphi$*$=\left(0,0,0,0,0,0,0,0,0\right)$.
Inverting \eqref{k} one can build the physical mass matrix as:
\begin{equation}
	\left(m^2\right)_i^{\ j}=K^{jk}\, \partial_k\partial_iV,
\end{equation}
with V the scalar potential.\\

\noindent
In the case at hand the scalar matrix at the origin is: 
\begin{equation}
	m^2|_{\varphi*}=\left(
	\begin{array}{ccccccccc}
		 \alpha ^2/2 & - \alpha ^2/2 & 0 & 0 & 0 & 0 & 0 & 0 & 0 \\
		- \alpha ^2/2 &  \alpha ^2/2 & 0 & 0 & 0 & 0 & 0 & 0 & 0 \\
		 \beta ^2/2 &  \beta ^2/2 &  \beta ^2 & 0 & 0 & 0 & 0 & 0 & 0 \\
		0 & 0 & 0 & \alpha ^2 & 0 & 0 & 0 & 0 & 0 \\
		0 & 0 & 0 & 0 & 0 & 0 & 0 & 0 & 0 \\
		0 & 0 & 0 & 0 & 0 & 0 & 0 & 0 & 0 \\
		0 & 0 & 0 & 0 & 0 & 0 & 0 & 0 & 0 \\
		0 & 0 & 0 & 0 & 0 & 0 & 0 & 0 & 0 \\
		0 & 0 & 0 & 0 & 0 & 0 & 0 & 0 &  \beta ^2 \\
	\end{array}
	\right).
\end{equation}
This can be easily diagonalized finding the following combination of fields to be massive:
\begin{equation}
	\begin{cases}
		\frac{1}{2}\left(\phi_1-\phi_2\right), \,\chi_1, \ \ \qquad\qquad m^2=\alpha^2,
		\\
		\frac{1}{2}\left(\phi_1+\phi_2+2\phi_3\right),\, \chi_6, \qquad m^2=\beta^2.
	\end{cases}
\end{equation}
Clearly to asses the actual scalar spectrum one also has to take into account the contribution of the covariant coupling \eqref{covder} at the origin, to check if some scalars get eaten by vectors to become massive. In our case we get the following contributions:
\begin{equation}
	\begin{split}
		{\cal L} \supset	-\frac{1}{2}\ \frac{g^2}{2^3}
			&\left[	\ \;\left|\frac{4}{g}\, \partial \chi_2 + \alpha A^{13}-\beta A^{24}\right|^2+
            \left|\frac{4}{g}\, \partial \chi_5+ \beta A^{13}-\alpha A^{24}\right|^2
			\right.
			\\
			&\left.+
			\left|\frac{4}{g}\, \partial \chi_4 + \alpha A^{14}+\beta A^{23}\right|^2 +\left|\frac{4}{g}\, \partial \chi_3 + \beta A^{14}+\alpha A^{23}\right|^2
			\right].
	\end{split}
\end{equation}
Clearly for generic $\alpha$ and $\beta$ the four combinations of vectors are all independent and one can choose a gauge in which all the four scalars present are eaten by the vectors, yielding four massive fields, two with $m^2=(\frac{\alpha+\beta}{2})^2$, and two with $m^2=(\frac{\alpha-\beta}{2})^2$. This gives the spectrum of Table \ref{tab:spectra}.
Therefore the massless spectrum is composed by the metric tensor, a two-form field, the dilaton, two vectors and the remaining orthogonal combination of $\phi_1,\phi_2, \phi_3$ with respect to $K_{ij}$.

There are also limiting cases, as for example $\alpha=0$ or $\beta=0$ where one has two extra massless scalars.
However, the more interesting cases are $\beta=\pm\alpha$. With these configurations there are  only two independent vectors among the previous four, and therefore  two more massless vectors and scalars. On top of this there are  also massless fermions. For $\beta=-\alpha$ there is a spin 1/2 fermion, while for $\beta=\alpha$ the fermion is accompanied by a gravitino, and thus the vacuum is supersymmetric. For this choice of gauging the spectrum organizes in a supergravity multiplet and a vector multiplet.
We will see in section \ref{sec:landscape_swampland} how these three very different configurations can all be constructed directly in string theory by means of different string orbifold constructions.

	\section{Generalized Scherk-Schwarz reductions}
\label{sec:DFT}

In this section we very briefly review the so-called Generalized Scherk-Schwarz (GSS) reductions in double field theory 
\cite{hull_non-geometric_2009,reid-edwards_flux_2009,aldazabal_effective_2011,grana_gauged_2012}
 (for a review see 
 \cite{aldazabal_double_2013}) or in generalized geometry \cite{lee_spheres_2017}, that lead from ten-dimensional supergravity to lower-dimensional gauged supergravity. We then show how the fluxes in each gauge orbit can be obtained in this form. We will concentrate on reductions to seven dimensions with O(3,3) duality group.

\subsection{Procedure}
In $T^3$ compactifications of the heterotic theory with frozen Wilson lines, or equivalently type II/I truncated to the common gravitational sector, the moduli can be encoded into a generalized $O(3,3)$ vielbein:
\begin{equation}
   E= \begin{pmatrix}
     e & b \,  \hat e \\
     0 & \hat e
    \label{E}
    \end{pmatrix}  ,
\end{equation}
as given in section \ref{sec:GSUGRA}. Clearly For reductions of type I, the field b is not the B-field but $C_2$.
In standard toroidal compactifications, these  do not depend on the internal space and become massless fields in the seven-dimensional supergravity theory. In generalized Scherk-Schwarz reductions the fields, or more precisely the generalized vielbein, depends on the internal coordinates in such a way that their (generalized) Lie algebra gives rise to a tensor with constant values whose components are (generalized) structure constants. This constant tensor is precisely the embedding tensor of the gauged supergravity. We quickly review the procedure.
The dependence on the internal coordinates $y$ (we suppress indices for the moment here) is fully encoded 
in the so-called duality twist matrix $U$, which should be an element of the duality group $G$, in our case O(3,3)
\begin{equation}
	U_A{}^{B}\left(y\right) \in G.
\end{equation}
\noindent
The ansatz for the reduction of the vielbein is:
\begin{equation} \label{Exy}
    \tilde E(x,y)_A{}^{\bar A}=U_A{}^{ B}(y) E_B{}^{\bar A}(x). 
\end{equation}
\noindent
The dependence on the internal coordinates should be such that the generalized Lie bracket of the vielbein is 
\begin{equation}
    [[U_A,U_B]]=f_{AB}{}^C\, U_C \, ,
\end{equation}
with constant $f_{AB}{}^C$. These "generalised fluxes" are those introduced in \eqref{fQ}, and give the embedding tensor of the gauged supergravity (here given by $\Q, \tilde{\Q} $).

Before introducing the generalized Lie bracket, we should underline the distinction between generalized geometry and double field theory. Both are formulations that incorporate the duality group of the theory using e.g. a generalised veilbein encoding the metric and the B-field. The difference lies in how this covariance extends to the internal space itself. While in generalised geometry the internal space is, for the case at hand, three-dimensional, and thus derivatives are not promoted to O(3,3) vectors $\partial_A$, in double field theory they are. The introduction of dual coordinates, or coordinates associated to windings, can be in principle be done for toroidal spaces, while in more involved geometries there is not always a clear meaning for this. Furthermore, consistency of the theory requires constraints on the dependence on the internal coordinates. There are two sufficient conditions for the consistency of the kind of reductions: the weak and the strong constraint. These two constraints are not necessary for the consistency in principle.
For what concerns us, we do not need the general expression for the generalized Lie bracket, but only how the O(3,3) structure constants are obtained from the twist matrix $U$ : 
\begin{equation} \label{fABC}
		f_{ABC}= 3 \ \eta_{_{D\left[A\right.}}\left(U^{-1}\right)^M_{\ \ B}\left(U^{-1}\right)^N_{\ \ C\left. \right]} \partial_M U^D_{\ \ N} . 
\end{equation}

Given the splitting \eqref{Exy} (where, again, the dependency on $y$ should be such that the structure constants \eqref{fABC} are actually constant, and furthermore that they satisfy Jacobi identities \eqref{Jacobi}), the reduced action is precisely the gauged O(3,3) action of the previous section, where
\begin{equation}
    \mathcal{M}_{AB}(x)=E_A^{\ \bar A}(x)\,  E_B^{\ \bar B} (x) \, \delta_{\bar A \bar B} \ ,
\end{equation}
and the dependence on the internal coordinates is fully encoded in the structure constants, giving rise to gauge covariant derivatives \eqref{covder} and the potential \eqref{gauged_pot}. 

One can parameterise the twist matrix $U \in O(3,3)/(O(3) \times O(3))$,  by a $GL(3)/SO(3)$ vielbein $u$ and a two-form $b$ as follows
\begin{equation}
     U^A{}_M = \left(\begin{matrix} u_a{}^m & u_a{}^n b_{nm} \\ 
0 & u^a{}_m \end{matrix}\right)\, . \label{TwistParam1}
\end{equation}
If these fields do not depend on dual coordinates, when computing the fluxes \eqref{fABC}, the only non-vanishing components are $H$ and $\omega$, while allowing dependence on dual coordinates introduces the flux $Q$. On the other hand, one can equivalently parameterise the twist matrix using a two-vector $\beta$ instead of $b$, as
\begin{equation}
     U^A{}_M = \left(\begin{matrix} u_a{}^m & 0 \\ u^a{}_n
\beta^{nm} & u^a{}_m \end{matrix}\right)\, .\label{TwistParam2}
\end{equation}
For this parameterisation, still not introducing dependence on the dual coordinates, one finds instead the fluxes $\omega, Q $ and $R$. For a general (over) parameterisation in terms of $b$ and $\beta$, and dependence on coordinates and dual coordinates, the fluxes $H, \omega, Q, R$ as derivatives of these fields were computed in \cite{aldazabal_double_2013}, we give them in \eqref{fluxes}. 

Clearly not all the flux configurations are possible due to the quadratic constraints required by closure of the algebra. For half-maximal supergravities it turns out  that these are weaker than the strong and weak constraints \cite{aldazabal_double_2013}.

    \section{String theory origin of fluxes}
\label{sec:GaugingsST}

We will first review very briefly non-geometric fluxes \cite{shelton_nongeometric_2005,Kachru:2002sk,Hellerman:2002ax} (for reviews see \cite{Wecht:2007wu,plauschinn_non-geometric_2019}). 
Then we analyze which vacua posses a known string theory realization and which not.

\subsection{Non-geometric fluxes}

Non-geometric fluxes arise naturally from T-duality acting on flux backgrounds. Consider $T^3$ with coordinates $x^1,x^2,x^3$ and $N$ units of $H$-flux,
\begin{equation}
	\int_{T^3} H = N,  \qquad ds^2 = (dx^1)^2+(dx^2)^2+(dx^3)^2 \, .
\end{equation}
One can choose a gauge in which
\begin{equation}
     B_{23}=Nx^1 \, .
\end{equation}
Buscher T-duality along $x^2$ trades the $H$-flux for a geometric (metric) flux $\omega_{31}{}^2=N$, and the background becomes the twisted torus
\begin{equation}
	ds^2 = \left(dx^2 - N x^1\, dx^3\right)^2 + (dx^3)^2+(dx^1)^2, \qquad B=0 \, .
\end{equation}
This is a circle fibration, with fiber coordinate $x^2$, over a $T^2$ base spanned by $(x^1,x^3)$, with monodromy $x^2\to x^2+Nx^1$. This is locally an ordinary Riemannian manifold, and still globally well-defined as a compact space (a nilmanifold).
The vielbein
\begin{equation}
    u^1=dx^1, \qquad u^2=dx^2-N x^1 \, dx^3 , \qquad u^3=dx^3,
\end{equation}
satisfies
\begin{equation}
    du^1=0, \qquad du^2 = N\, u^3\wedge u^1, \qquad du^3=0 \ .
\end{equation}
The vector fields $Z_1,Z_2,Z_3$ generating left translations on this group manifold, defined to be dual to the left-invariant one-forms, satisfy
\begin{equation}
	[Z_3,Z_1]=N\,Z_2\,, \qquad [Z_1,Z_2]=0\,, \qquad [Z_2,Z_3]=0\,,
\end{equation}
consistent with this background, according to \eqref{commutators}.

A further T-duality along $x^3$ -- one of the two base directions -- singles out $x^1$ as the remaining base circle, over which $(x^2,x^3)$ now form the fibered $T^2$. This second duality produces the metric and $B$-field
\begin{equation}
	ds^2 = \frac{1}{1+N^2(x^1)^2}\left((dx^2)^2+(dx^3)^2\right)+(dx^1)^2\,, \qquad B_{23}=\frac{Nx^1}{1+N^2(x^1)^2}\, .
\end{equation}
To understand better what is happening, consider the Kähler modulus of this fibered $T^2$, $\rho=B_{23}+i\sqrt{g}$. Going once around the base coordinate $x^1$, it transforms as $1/\rho\to 1/\rho+N$. The background is therefore only locally geometric: patching together the local charts around $x^1$ requires T-duality transition functions rather than diffeomorphisms plus gauge transformations. This is a $T$-fold \cite{Hull:2006va} with non-geometric flux
\begin{equation}
	 Q_1{}^{23}=N\, .
\end{equation}
This situation can also be described using the bi-vector field $\beta$ introduced in Section~\ref{sec:DFT}. Under T-duality along $x^2$ and $x^3$, the $B$-field component $B_{23}$ is mapped to the bi-vector component $\beta^{23}$, with  a linear dependence on $x^1$. The appearance of $Q$-flux can then also be understood as a monodromy of this bi-vector.\footnote{Note, though, that the local gauge transformations of the $B$-field, which undergoes a global monodromy realised by $H$-flux, are described in mathematical terms by the concept of gerbes, but there is no analogous structure for a bi-vector.}

A third, purely formal T-duality along $x^1$, which is not even an isometry of the background, and is therefore not a legitimate duality transformation, would formally produce the $R$-flux, for which not even a local geometric description survives. Altogether one has the chain
\begin{equation}
	H_{123} \;\xrightarrow{T_2}\; \omega_{31}{}^2 \;\xrightarrow{T_3}\; Q_1{}^{23} \;\xrightarrow{T_1}\; R^{123}\, .
\end{equation}
Keeping all four flux classes on the same footing is precisely what restores full T-duality covariance of the lower-dimensional scalar potential \cite{plauschinn_non-geometric_2019}.

Note, however, that the presence of $Q$- and $R$-flux does not necessarily mean that the corresponding background is non-geometric. As we saw in the previous section, a generalised Scherk-Schwarz reduction with all fields independent of the coordinates, but parameterised by a vielbein and a bi-vector field $\beta$, can generate $Q$- and even $R$-flux. The distinction between geometric and non-geometric flux backgrounds is therefore not always straightforward to draw from the fluxes alone.

\subsection{Fluxes and O(3,3) gaugings}

The possible orbits of gaugings with $H,\omega,Q,R$ fluxes  are given in Table \ref{tab:gaugings}. In what follows we inspect each entry of table \ref{tab:gaugings} and argue if the gauging and possible vacua come from a geometric compactification or not, and in the former we give the string theory origin. Most of these gaugings were identified already in \cite{dibitetto_duality_2012}. Here we do a systematic analysis.
We go in increasing order of complexity, from 11 to 1 (and for pedagogical reasons, not completely in order):

\begin{itemize}
	\item [11:] This gauging comes from a geometric compactification with $\alpha$ units of $H$-flux on the group manifold specified by the Heisenberg algebra:
	\begin{equation}
		\left[Z_1,Z_2\right]=\beta \;Z_3, \quad 	\left[Z_3,Z_1\right]=0, \quad	\left[Z_2,Z_3\right]=0. \nn
	\end{equation}
	This group manifold is non-compact. One can make it compact via identifications of the coordinates, or in other words by finding a lattice that is preserved under the action of the group. This is only possible if $\beta \in {\mathbb Z}$. Furthermore, the total $H$-flux is quantised, requiring $\alpha \in {\mathbb Z}$.\footnote{Here we work in units $(2\pi)^2\alpha'=1$.} Both the $H$-flux and the (negative) curvature of the Heisenberg manifold contribute positively to the potential (in a runaway fashion), so there is no vacuum supported by these fluxes alone.

	\item [10:] This configuration is clearer after dualizing along direction 3. One now has $\beta$ units of $H$-flux on a group manifold with the ISO(1,1) algebra:
	\begin{equation}
		\left[Z_1,Z_2\right]=\alpha \;Z_3, \quad 	\left[Z_3,Z_1\right]=-\alpha \;Z_2, \quad	\left[Z_2,Z_3\right]=0. \nn
	\end{equation}
	This group manifold is compactifiable if $2\cosh{\alpha}=n$, with $n \in {\mathbf N}$, $n>2$. Nevertheless this configuration does not admit a vacuum, for the same reason as configuration 11 above.

	\item [8:] After dualizing along direction 3, these fluxes likewise correspond to $\beta$ units of $H$-flux on a group manifold, this time with ISO(2) algebra,
    \begin{equation}
		\left[Z_1,Z_2\right]= \alpha \;Z_3, \quad 	\left[Z_3,Z_1\right]= \alpha \;Z_2, \quad	\left[Z_2,Z_3\right]=0. \nn
	\end{equation}
    ISO(2) is compactifiable for $\alpha= \frac{1}{N}$, with $N\in \mathbb{Z}$. 
    This gauging admits a vacuum only for $\beta=0$: the $H$-flux gives a positive contribution to the potential, while the ISO(2) group manifold is Ricci-flat and so does gives no contribution. 

\end{itemize} 
   \noindent The Heisenberg, ISO(1,1) and ISO(2) algebras are the only 3d unimodular algebras with an abelian ideal $\left[Z_2,Z_3\right]=0$. As group manifolds this realizes them as $T^2$ fibrations over $S^1$, with parabolic, hyperbolic and elliptic monodromy respectively.

\begin{itemize}

\item [5] This configuration can be dualized to a geometric one with $\beta$ units of $H$-flux on the group manifold with algebra:
	\begin{equation}
		\left[Z_1,Z_2\right]= \alpha \;Z_3, \quad 	\left[Z_3,Z_1\right]= \alpha \; Z_2, \quad	\left[Z_2,Z_3\right]=- \alpha\; Z_1. \nn
	\end{equation}
     Here one has the algebra $SO(2,1) \simeq SL(2,\mathbb{R})$, whose group manifold is compactifiable for any value of $\alpha$. Flux quantization implies $\beta/\alpha^3 \in \mathbb{Z}$.\footnote{We use unit-period coordinates for the group manifold directions, as in the previous cases; the flux integral is then $\int H = \beta/\alpha^3$, quantized to an integer.}
     The manifold is negatively curved, and both the curvature and the flux contribute positively to the potential, so there is no minimum.

\item [4] This is similar to the previous case, but with a key sign difference: the algebra is now
	\begin{equation}
		\left[Z_1,Z_2\right]= \alpha \;Z_3, \quad 	\left[Z_3,Z_1\right]= \alpha \; Z_2, \quad	\left[Z_2,Z_3\right]= \alpha\; Z_1. \nn
	\end{equation}
     In this case one finds an $SU(2)$ group manifold, which is itself compact. Flux quantization implies, as before, $\beta/\alpha^3 \in \mathbb{Z}$. The only possibility of a vacuum (up to the runaway dilaton discussed earlier) is $\beta=\alpha$. The curvature contributes negatively to the potential, while the flux contributes positively; since the two scale differently with $1/R$ (where $R=1/\alpha$ is the radius of the 3-sphere), the combination of both gives a minimum with negative potential, but only for $\beta=\alpha$.

\end{itemize}

We have now covered all the possible 3-dimensional Lie groups. The gaugings that follow are not dualizable to a geometric configuration. We will nevertheless see that some do still admit a geometric realisation.

\begin{itemize}
    \item [9] This is the first instance in our discussion where, for both $\alpha$ and $\beta$ non-zero, one cannot dualize away the non-geometric $Q$-flux. This gauging does not have a vacuum for any $\alpha$ or $\beta$, so it will not be realizable as an asymmetric orbifold. It is thus unclear whether this gauging has a string theory origin. 
    
    \item [7] As before, there is no frame in which the non-geometric fluxes are absent, and the potential does not have a minimum unless $\beta=0$. In that case one has the ISO(2) geometry without $H$-flux, so there is a vacuum as in configuration~8. It remains unclear whether this gauging has a known string theory origin whenever both $\alpha$ and $\beta$ are non-zero.
	
	\item [6] Here one has another case that cannot be dualized to any geometric configuration for generic $\alpha$ and $\beta$. However, this case has a vacuum with vanishing potential for any $\alpha$ and $\beta$. We will see that for certain quantised values of $\alpha$ and $\beta$, the fluxes correspond to an asymmetric orbifold configuration.

	\item [3,2] These two cases can never be dualized to a frame in which non-geometric fluxes are absent, not even when $\alpha$ or $\beta$ vanish, and there is no known string theory origin for these gaugings. Gauging 3 has minimum with vanishing potential for $\beta=-\alpha$, so this could in principle be realized by an asymmetric orbifold, but we will see that there is no such realisation.
    
    \item [1] Again, the non-geometric fluxes cannot be dualized away. However, for $\beta=0$ or $\beta=-\alpha$ this does have a geometric realization: it comes from the reduction over $S^3$ with $\alpha$ units of $H$-flux \cite{lee_spheres_2017}. This reduction gives a different gauging than configuration~4 because it requires a Generalized Scherk-Schwarz reduction rather than an ordinary one, with different vielbeins for the left- and right-moving sectors, though with no dependence on dual coordinates. This configuration is then genuinely geometric, but the reduction is not the standard Scherk-Schwarz one. This implies that the modes kept in the (consistent) truncation differ from those kept in an ordinary Scherk-Schwarz reduction, which can be seen from the mass matrix of the scalars: they are all massless for gauging~1, while in gauging~4 they are massive (or eaten by a vector to become massive). This is a nice counterexample to the simplified reasoning that ``a configuration with $Q$- or $R$-flux that cannot be T-dualized to one without it is non-geometric.'' 

    Note, however, that for these cases ($\beta=0$ or $\beta=-\alpha$) the potential at the origin is non-zero, so these are not really vacua, the dilaton is runaway, as explained before. This internal $S^3$ geometry could be part of a true vacuum of the form $AdS_3\times T^4\times S^3$ with $H$-flux on both the $S^3$ and the $AdS_3$ part, or by coupling it to a linear dilaton solution as in the near-horizon geometry of NS5-branes. \\
    On the other hand, for $\beta\neq 0$ and $\beta\neq -\alpha$ there is no known string theory origin. For $\beta=\alpha$ the potential is zero, so this configuration could, as in case~3, also potentially be realized as an asymmetric orbifold. We will see, however, that there is no such realisation.

\end{itemize}
	
	\section{Orbifolds and \texorpdfstring{$O(3,3)$}{} gauged supergravities}

\label{sec:vacuaCFT}
In this section we present the tools needed to build exact world-sheet descriptions of some of the gauged sugravity vacua presented in section \ref{sec:GSUGRA}. We first  review briefly the machinery of orbifold constructions \cite{dixon_strings_1985,dixon_strings_1986} of type II theories which will be our main tool. We then move to discuss the connection between a given gauged supergravity vacuum and its possible orbifold realization, leaving the explicit realization of some of the gauged supergravity vacua to the next section.

\subsection{Orbifold construction}
The orbifold procedure of a string theory model with an exact CFT description gives  another model with an exact CFT description.
Orbifolding a theory consists in gauging a discrete symmetry of the underlying CFT. Given a group $\mathds{G}$ of discrete automorphisms of the CFT,  one first projects onto the states of the Hilbert space $\mathcal{H}$ that are invariant under the action of $\mathds{G}$:
\begin{equation}
    \hat{\mathcal{G}} \ket{\phi}=\ket{\phi} , \quad \forall\, \hat{\mathcal{G}}\in\text{Rep}(\mathds{G},\mathcal{H}). 
\end{equation}
The above projection brakes the modular invariance of the original theory. To restore this crucial property one has to include the so called twisted sectors, which are new sectors of the theory where the strings are closed up to the action of the group one is gauging:
\begin{equation}
    X(\tau,\sigma+2\pi)=\mathcal{G}X(\tau,\sigma), \quad \forall \mathcal{G}\in\mathds{G}.
\end{equation}
The above equation defines string states that live in the so called twisted Hilbert spaces $\mathcal{H}_{\mathcal{G}}$, that once projected form the full
spectrum of the orbifolded theory. A fundamental tool to deal with orbifold is the one-loop partition function, which is defined as the following trace over the Hilbert space:
\begin{equation}
    \mathcal{Z}(\tau,\bar{\tau})=\text{Tr}_{\mathcal{H}}\left(q^{L_0-\frac{c}{24}}\bar{q}^{\bar{L}_0-\frac{c}{24}}\right),
\end{equation}
where $L_0$ is the Virasoro zero mode, $c$ the central charge and $q=e^{2i\pi \tau}$, with $\tau$ the modular parameter of the worldsheet torus. The partition function counts the, spin statistic signed, number of states at each mass level of the theory.

For the sake of clarity let us consider the class of abelian orbifolds $\mathds{G}=\mathbb{Z}_N$. In this particular case the group $\mathds{G}$ is generated by only one element $\mathcal{G}$, all the others are powers of this generator. For each sector twisted by $\mathcal{G}^h$ and projected by $\mathcal{G}^g$ one has a partial trace of the form:
\begin{equation}
    \mathcal{Z}\left[\genfrac{}{}{0pt}{}{h}{g}\right](\tau,\bar{\tau})=\text{Tr}_{\mathcal{H}_{\mathcal{G}^h}}\left(\hat{\mathcal{G}}^gq^{L_0-\frac{c}{24}}\bar{q}^{\bar{L}_0-\frac{c}{24}}\right).
    \label{eq:PFunc}
\end{equation}
The partition function of the orbifold is given by summing over $g,h=0,...,N-1$ and dividig by N:
\begin{equation}
    \mathcal{Z}^{\text{orb}}=\frac{1}{N}\sum_{h,g\in \mathbb{Z}_N} \mathcal{Z}\left[\genfrac{}{}{0pt}{}{h}{g}\right].
\end{equation}
Naturally not all the discrete symmetries of the original theory can be gauged due to the possible presence of anomalies. The absence of (one loop) anomaly for an orbifold is literally the requirement of a modular-invariant partition function.
In this formalism modular invariance is implied by the modular covariance of the blocks \eqref{eq:PFunc}:
\begin{equation}
\mathcal{Z}\left[\genfrac{}{}{0pt}{}{h}{g}\right]\left(\frac{a \tau+b}{c\tau+d},\frac{a\bar{\tau}+b}{c\bar{\tau}+d}\right)=\mathcal{Z}\left[\genfrac{}{}{0pt}{}{ah+cg}{dg+bh}\right](\tau,\bar{\tau}),\quad \forall 
\begin{pmatrix}
    a&b\\
    c&d
\end{pmatrix}\in SL(2,\mathbb{Z}),
\end{equation}
and the N-periodicity of both the characteristics:
\begin{equation}
    \mathcal{Z}\left[\genfrac{}{}{0pt}{}{h+N}{g}\right]= \mathcal{Z}\left[\genfrac{}{}{0pt}{}{h}{g}\right],
    \qquad
     \mathcal{Z}\left[\genfrac{}{}{0pt}{}{h}{g+N}\right]= \mathcal{Z}\left[\genfrac{}{}{0pt}{}{h}{g}\right].
\end{equation}
It has been shown in 
\cite{narain_asymmetric_1987,narain_asymmetric_1991} that in the context of abelian orbifolds level-matching is not just necessary, but it is also sufficient for modular invariance and hence for the absence of anomaly. The level-matching condition has a concise form that we write below in \eqref{eq:ModInv}.

Let us now enter a bit more in the details of the orbifold constructions we are interested in. Consider type II string theory compactified on $T^d$. The states of the theory, among other quantum numbers, are labeled by the Narain momenta $(P_L;P_R)$ which live in the Narain lattice $\Gamma_{d,d}$, and can be expressed in terms of the string momenta and windings along the compactified directions. Each momentum $P_L / P_R$ is conjugate to a chiral coordinate $X_L / X_R$, which in turn is related by worldsheet supersymmetry to a chiral Majorana worldsheet fermion $\psi_L / \psi_R$. Given this, to gauge a symmetry of the Narain lattice\footnote{We defer the discussion on the symmetries of the Narain lattice to appendix \ref{App:Narain_Sym}.} one has to make sure  that the orbifold action acts consistently on all the related degrees of freedom. 
A necessary condition for a symmetry of the Narain lattice to uplift to a full symmetry of the string  CFT is that it preserves the left-right splitting of the Hilber space. That said, let us consider a symmetry of the Narain lattice that acts on the momenta as:
\begin{equation}
    \hat{\mathcal{G}}\ket{P_L,P_R}=e^{2i\pi\left(P_L\cdot v_L-P_R\cdot v_R\right)}\ket{\theta_LP_L,\theta_RP_R},
    \label{eq:act}
\end{equation}
where, in general there is a $\theta=\text{diag}(\theta_L,\theta_R)$, with $\theta_{L(R)} \in SO(d)$  and a vector $v=(v_L,v_R)$, associated to each element $\mathcal{G}\in\mathds{G}$. 
The twist part of the action in \eqref{eq:act} has to be done also on the chiral coordinates in order to preserve the wave operator $e^{i \left(P_L\cdot X_L-P_R \cdot X_R\right)}$,
and consequently on the worldsheet fermions in order to preserve the worldsheet supercurrent $\partial X_L \cdot \psi_L+\bar{\partial}X_R\cdot \psi_R$. It follows that the action on the Narain momenta is accompanied by:
\begin{equation}
    \mathcal{G}X_{L,R}=\theta_{L,R} \ X_{L,R},
    \qquad
    \mathcal{G}\psi_{L,R}=\theta_{L,R}\ \psi_{L,R}.
\end{equation}
When $\theta_L \neq \theta_R$ or $v_L \neq v_R$ the orbifold is said to be asymmetric. Let us now consider the most general $\theta$ for $\mathds{G}=\mathbb{Z}_N$ in a diagonal basis:
\begin{equation}
   \begin{split}
    \theta&=
    \begin{pmatrix}
        \theta_L &0\\
        0&\theta_R
    \end{pmatrix} 
     \\
    &=\text{diag}(e^{2i\pi \lambda_L^1/N},e^{-2i\pi \lambda_L^1/N},...,e^{-2i\pi \lambda_L^{{\left\lfloor {d}/{2} \right\rfloor}}/N},\pm1
    , \ e^{2i\pi \lambda_R^1/N},e^{-2i\pi \lambda_R^1/N},...,e^{-2i\pi \lambda_R^{\left\lfloor {d}/{2} \right\rfloor}/N},\pm1),
    \end{split} \nonumber
\end{equation}
where the integer part and the $\pm$ are present only if d is odd. Notice that the structure of eigenvalues is dictated by the requirement that $\theta$ is a symmetry of the lattice (see appendix \ref{App:Narain_Sym}).
In order to have a consistent action, one needs
\begin{equation}
    \lambda_{L,R}^i\in \mathbb{Z} \ , \qquad N\, v \in \Gamma_{d,d}
\end{equation}
Moreover, a consistent uplift of this action to the fermions requires $\sum_{i}\lambda^i_{L,R}\in 2\, \mathbb{Z}$. If this is not satisfied, then the order of the orbifold action is $2N$ rather than $N$.
In this specific setting, level matching can be stated in the very concise form \cite{baykara_quasicrystalline_2025}:
\begin{equation}
    \sum_{i} \left(\left\{\frac{\lambda^i_L}{N}\right\}-\left\{\frac{\lambda^i_R}{N}\right\}\right)+(v_L^*)^2-(v_R^*)^2
    \in\frac{2\mathbb{Z}}{\hat{N}} \ ,
    \label{eq:ModInv}
\end{equation}
with
\begin{equation}
     \hat{N}=\begin{cases}
        N, \ \ \ {\rm for} \ \left(\sum_{i}\lambda^i_{L}\right) \wedge \left(\sum_{i}\lambda^i_{R}\right)\in 2\mathbb{Z},
        \\
        2N, \,  \ {\rm otherwise}. \ 
         \ 
        \label{hatN}
    \end{cases}
    \end{equation}

\noindent
Here $v^*$ is the projection of the shift vector $v$ onto the invariant lattice $I$, defined as $I=\{P\in\Gamma_{d,d}\,|\, \theta P=P \}$, and $\{\cdot\}$ is the fractional part. In the case of a $\mathbb{Z}_N\times \mathbb{Z}_M$ orbifold one has a condition of the form \eqref{eq:ModInv} for each independent generator, with appropriate order.

\subsection{ \texorpdfstring{$T^2 \times S^1$}{} freely acting orbifolds}

In this section we specialize to abelian orbifolds of $T^2\times S^1$, with some twist on $T^2$ accompanied by a shift along $S^1$, such that the orbifold action act freely on the coordinates.
We first analyze the allowed abelian twists on $T^2$ and then the choice of shifts.

\subsubsection*{Allowed orbifold actions on   \texorpdfstring{$T^2$}{}}

The Abelian discrete symmetries of the  Narain lattice $\Gamma_{2,2}$ are detailed in Appendix \ref{App:Narain_Sym}. Some of these symmetries are present only at special points in the moduli space, which we parameterize as usual in terms of Kahler and complex structure moduli $T=T_1+iT_2$, $U=U_1+iU_2$ as follows:
\begin{equation}
        G=\frac{T_2}{U_2}\begin{pmatrix}
            1&U_1\\
            U_1& |U|^2
        \end{pmatrix},
        \ \quad 
        B=T_1\begin{pmatrix}
            0&1\\
            -1&0
        \end{pmatrix}.
\end{equation}
The only rotational symmetries of the $T^2$ Narain lattice are given, up to T-duality, by the following rotation parameters, and fix the moduli to:

\begin{table}[H]
\adjustbox{width=0.35\textwidth}{
    \centering
    \renewcommand{\arraystretch}{1.2}
    \begin{tabular}{|c|c|c|c|c|}
        \hline
        $\left(\frac{\lambda_L}{N};\frac{\lambda_R}{N}\right)$&$T_1$&$T_2$&$U_1$&$U_2$ \\
        \hline\hline
     ($\frac{1}{2}$;$\frac{1}{2}$)& -& -& -&-\\
     \hline
     ($\frac{1}{3}$;$\frac{1}{3}$)& -& -& $\frac{1}{2}$&$\frac{\sqrt{3}}{2}$ \\
          \hline
     ($\frac{2}{3}$;$\frac{2}{3}$)& -& -& $\frac{1}{2}$&$\frac{\sqrt{3}}{2}$ \\
     \hline
     ($\frac{1}{4}$;$\frac{1}{4}$)& -& -& 0&1 \\
     \hline
     ($\frac{1}{6}$;$\frac{1}{6}$)& -& -& $\frac{1}{2}$&$\frac{\sqrt{3}}{2}$ \\
     \hline
     ($\frac{1}{2}$;0)& 0& 1& 0&1 \\
     \hline
     ($\frac{1}{3};0$) &$\frac{1}{2}$&$\frac{\sqrt{3}}{2}$& $\frac{1}{2}$&$\frac{\sqrt{3}}{2}$ \\
     \hline
     ($\frac{2}{3};0$) &$\frac{1}{2}$&$\frac{\sqrt{3}}{2}$& $\frac{1}{2}$&$\frac{\sqrt{3}}{2}$ \\
     \hline
     ($\frac{1}{12}$;$\frac{5 \ }{12}$)&$\frac{1}{2}$&$\frac{\sqrt{3}}{2}$& 0&1 \\
     \hline
    \end{tabular}
    \caption{Possible rotational symmetries of the Narain lattice of $T^2$ up to T-duality. A - indicates that the moduli are free.}
    \label{tab:rotsym}
    }
\end{table}
\noindent 
The first five lines correspond to symmetric orbifold actions, while the last  four give rise to asymmetric orbifolds. In particular, the actions in the sixth, seventh and eighth lines are purely holomorphic crystallographic actions, while the last line is the only non crystallographic action, corresponding to
a rotation that is a symmetry of the Narain lattice but not of the underlying geometric
$T^2$ lattice itself, since it forces the left- and right-moving momenta to take irrational
values in the natural lattice basis. 
In a freely acting orbifold these rotational actions are accompanied by shifts on the circle direction.

\subsubsection*{The full orbifold action}

We denote the coordinates on $T^3$ by $x^a$, $a=1,2,3$, and split them into an $S^1$ coordinate $x^1\equiv y$ and a complex coordinate on $T^2$: $z=x^2+ix^3$. We further split the corresponding fields into left and right moving components. The orbifolds we implement are freely acting $\mathbb{Z}_{2N}$ orbifolds with the following generator\footnote{As we will explain in detail in section \ref{sec:explicitOrb}, the lift of the bosonic action to the fermions will force us, in most of the cases, to consider an order $2N$ orbifold. To avoid level-matched tachyons for generic values of the moduli we therefore employ an order $2N$ shift.}

   \begin{equation}
  G:\quad
  z_{L(R)}\;\longrightarrow\; e^{2i\pi\lambda_{L(R)}/N}\,z_{L(R)},
  \qquad
  y_{L(R)}\;\longrightarrow\; y_{L(R)}-\frac{\pi R}{2N}.
  \label{eq:orbifold_generators}
\end{equation}
In principle one can do a $\ZZ_{2N} \times \ZZ_{2M}$ orbifold, with different shifts for each action. A generic shift can be decomposed into a symmetric, "momentum shift", of order $2N$, as in 
\eqref{eq:orbifold_generators}, and an asymmetric, "winding" shift of the form 
$y_{L(R)}\;\longrightarrow\; y_{L(R)} \pm \frac{\pi }{2MR}$. This choice is severely constrained by modular invariance. Indeed, for $T^3$ we found no modular action with multiple $\ZZ_N$ factors. Likewise, on $T^3$ there are no modular actions with $M\neq N$ and pure momentum or pure winding shift on the generators. For more generic shifts, one can have modular invariant actions with multiple  factors and $N\neq M$. However, these turn out to realize the same gaugings as a single shift.  
We then restrict then to a single $\ZZ_{2N}$ with a pure momentum shift, as in \eqref{eq:orbifold_generators}.

\subsection{Link between orbifolds and gaugings}

We focus on vector states states carrying Narain momentum $P$ only along the $S^1$
direction, since these are the lightest vectors that acquire mass through the stringy
Scherk--Schwarz mechanism \cite{scherk_spontaneous_1979,rohm_spontaneous_1984,kounnas_spontaneous_1988}.
 There are three left-moving and three
right-moving vectors, whose vertex operators in the $(-1)$-picture are
\begin{equation}
  V^a_{-1}
  = \psi^\mu\,\bar{\psi}^a\,e^{ik\cdot X}\,e^{iq^a_+y_L}\,e^{iq^a_-y_R}\,
    e^{-\varphi}e^{-\bar\varphi},
  \qquad
  V^{a'}_{-1}
  = \psi^{a'}\,\bar\psi^\mu\,e^{ik\cdot X}\,e^{iq'^{a'}_+y_L}\,
    e^{iq'^{a'}_-y_R}\,e^{-\varphi}e^{-\bar\varphi},
  \label{eq:vertex_operators_m1}
\end{equation}
(no sum on $a$ and $a'$), with\footnote{For clarity, we we use $q$ instead of $P$ to denote momenta along the circle direction, and subindices $\pm$ for its left and right components.}
\begin{equation}
    q^a_\pm=\frac{m^a}{R}\pm n^aR \, \in\Gamma^{1,1}(R)
    \label{qpm}
\end{equation}
    and
similarly for $q^{\prime a}_\pm$. 
For simplicity we derive the
conditions for $V$; the analogous result for $V'$ follows by
$\lambda_R\leftrightarrow\lambda_L$.

Under the generator ${\cal G}$, the complex worldsheet fermion  is rotated
as 
\begin{equation}
 \bar\psi^a\to e^{2i\pi\lambda^{(a)}_R/N}\bar\psi^a \, ,  
\end{equation}
where
$\lambda^{(a)}_{R}=(0,\lambda_R,-\lambda_R)$ are the $T^2$-rotation
eigenvalues in the basis $a=1,2,3$.  The momentum exponentials are shifted as
\begin{equation}
e^{iq^a_\pm y_{L/R}}\to e^{iq^a_\pm y_{L/R}} e^{-i\pi q^a_\pm R/2N}\, ,    
\end{equation}
where $\pm$ is correlated with $L(R)$.
Invariance of $V^a_{-1}$ under ${\cal G}$  requires the total phase
to vanish, i.e.
\begin{equation}
  \lambda^{(a)}_{R}
  = \tfrac{1}{4}q^a_+R+\tfrac{1}{4}q^a_-R
  \pmod{N}.
  \label{eq:inv}
\end{equation}
To focus on the lowest-mass states, we solve~\eqref{eq:inv} as a strict
equality (not merely mod~$N$), which uniquely fixes $m^a$ and $n^a$ for
a given set of shifts.
Explicitely
\begin{align}
    \lambda^{(a)}_{R}
  = \tfrac{R}{4}(q^a_++q^a_-)
   = \frac{m^a}{2},
\label{eq:inv2}
\end{align}
where we used \eqref{qpm}.
This condition gives
\begin{equation}
  q^a_\pm
  = \frac{2\lambda^{(a)}_{R}}{R}\pm n^aR.
  \label{eq:qa_pm_solution}
\end{equation}
The level-matching condition requires
\begin{equation}
\frac12 \left(   (q^a_+)^2-(q^a_-)^2 \right)
  = 2\,m^a n^a 
  = 2\,(2\lambda^{(a)}_R) \,n^a =0  \quad \forall a.
  \label{lm}
\end{equation}
This equation can be satisfied by $n^2=n^3=0$ and any $n^1$, but since we are interested in the lowest mass states, we set all  $n^a = 0$. The momenta along the circle for each vector $V^a$, $V^{a'}$ is thus given by
\begin{equation}
  q^a_\pm
  = \left(0,\;
    \frac{2\lambda_R}{R},\;
    -\frac{2\lambda_R}{R}
    \right),
    \qquad
    q^{\prime a'}_\pm
  = \left(0,\;
    \frac{2\lambda_L}{R},\;
    -\frac{2\lambda_L}{R}
    \right).
  \label{eq:qa_pm_explicit}
\end{equation}

To read off the gauge algebra of the effective theory we compute the
tree-level three-point amplitudes of the vector bosons.  We need the
vertices in the $0$-picture, which in our conventions are
\begin{align}
  V^a_0
  &= 2\!\left(i\partial X^\mu+\tfrac{1}{2}(\psi\cdot\hat{K}_L)\psi^\mu\right)2
     \!\left(i\bar\partial X^a+\tfrac{1}{2}(\bar\psi\cdot\hat{K}_R)\bar\psi^a\right)
     e^{ik\cdot X}e^{iq^a_+y_L}e^{iq^a_-y_R},
     \\
  V^{a'}_0
  &= 2\!\left(i\partial X^{a'}+\tfrac{1}{2}(\psi\cdot\hat{K}'_L)\psi^{a'}\right)2
     \!\left(i\bar\partial X^\mu+\tfrac{1}{2}(\bar\psi\cdot\hat{K}'_R)\bar\psi^\mu\right)
     e^{ik\cdot X}e^{iq'^{a'}_+y_L}e^{iq'^{a'}_-y_R},
  \label{eq:V0aprime}
\end{align}
where normal ordering is understood and we have defined the total left/right momenta
\begin{equation}
  \hat{K}^\Sigma_{L(R)}
  = \bigl(k^\sigma,\,0,\,0,\,q^a_{\pm}\bigr),
  \qquad
  \hat{K}^{\prime\Sigma}_{L(R)}
  = \bigl(k^\sigma,\,0,\,0,\,q^{\prime a}_{\pm}\bigr),
  \label{eq:K_hat}
\end{equation}
with the first entries referring to spacetime and the last three to
internal momenta.
 
Using $\mathrm{SL}(2;\mathbb{C})$ invariance we fix the three vertex
positions to $0$, $z_\infty$, $1$, sending $z_\infty\to\infty$ at the
end.  The superghost background charge $+2$ is soaked up by choosing one
$0$-picture and two $(-1)$-picture insertions.  Denoting a generic vertex
by $V^A$ with $A\in\{a,a'\}$, the amplitudes we want to compute are:
\begin{equation}
  \mathcal{A}_{3\text{pt}}
  \sim
  \langle\!\langle
    V^A_0(0)\,V^B_{-1}(z_\infty)\,V^C_{-1}(1)
  \rangle\!\rangle_{\text{m}+\text{gh}}\!\! \ \Big|_{|z_\infty|\gg 1}
  = |z_\infty|^2\,
  \langle\!\langle
    V^A_0(0)\,V^B_{-1}(z_\infty)\,V^C_{-1}(1)
  \rangle\!\rangle_{\text{m}},
  \label{eq:3pt_amplitude}
\end{equation}
where the ghost factor $|z_\infty|^2$ cancels the large-$z_\infty$
behaviour of the matter correlator.
 
We compute the amplitude for three $V^a$ vertices in detail; the other
cases are analogous.  Expanding the $0$-picture vertex and using the
correlators in Appendix~A, we split the matter amplitude into four
contributions according to how the $\psi^\mu$ and $\bar\psi^a$
contractions are distributed:
\begin{align}
  \langle\!\langle
    V^a_0(0)\,V^b_{-1}(z_\infty)\,V^c_{-1}(1)
  \rangle\!\rangle
  &= I + II + II' + III,
  \label{eq:amplitude_split}
\end{align}
where
\begin{equation}
\begin{aligned}
  I
  &= -4\langle\psi^\nu(z_\infty)\psi^\rho(1)\rangle
      \langle\bar\psi^b(\bar z_\infty)\bar\psi^c(1)\rangle
      \langle{:}\partial X^\mu(0)V_X{:}\rangle
      \langle{:}\bar\partial x^a(0)V_y{:}\rangle,
    \\
  II
  &= 2i\hat{K}^{\Sigma}_L
     \langle{:}\psi_\Sigma\psi^\mu{:}(0)\,\psi^\nu(z_\infty)\,\psi^\rho(1)\rangle
     \langle\bar\psi^b(\bar z_\infty)\bar\psi^c(1)\rangle
     \langle V_X\rangle
     \langle{:}\bar\partial x^a(0)V_y{:}\rangle,
    \\
  II'
  &= 2i\hat{K}^{\Pi}_R
     \langle\psi^\nu(z_\infty)\psi^\rho(1)\rangle
     \langle{:}\bar\psi_\Pi\bar\psi^a{:}(0)\,\bar\psi^b(\bar z_\infty)\,
       \bar\psi^c(1)\rangle
     \langle{:}\partial X^\mu(0)V_X{:}\rangle
     \langle V_y\rangle,
    \\
  III
  &= \hat{K}^{\Sigma}_L\hat{K}^{\Pi}_R
     \langle{:}\psi_\Sigma\psi^\mu{:}(0)\,\psi^\nu(z_\infty)\,\psi^\rho(1)\rangle
     \langle{:}\bar\psi_\Pi\bar\psi^a{:}(0)\,\bar\psi^b(\bar z_\infty)\,
       \bar\psi^c(1)\rangle
     \langle V_X\rangle\langle V_y\rangle \, ,
\end{aligned}
\end{equation}
where we  defined
\begin{equation}
\begin{aligned}
    &V_X=e^{ik_1\cdot X(0,0)}e^{ik_2\cdot X(z_\infty,\bar z_\infty)}
    e^{ik_3\cdot X(1,1)},
    \\
    &V_y=e^{iq^a_+y_L(0)}e^{iq^b_+y_L(z_\infty)}e^{iq^c_+y_L(1)}
    e^{iq^a_-y_R(0)}e^{iq^b_-y_R(\bar z_\infty)}e^{iq^c_-y_R(1)}.
\end{aligned}
\end{equation}
Using the identities in Appendix \ref{App:conventions} and keeping only the leading terms for
$|z_\infty|\gg 1$ (which requires the vertices to be on-shell, namely to satisfy  \eqref{lm},
as well  conservation of both external and internal momenta, respectively  $k\equiv k_1+k_2+k_3=0$ and $q^a_\pm+q^b_\pm+q^c_\pm=0$), one finds:
\begin{equation}\label{eq:terms_explicit}
\begin{aligned}
  I   &\simeq \frac{\delta(k)}{|z_\infty|^2}
              \bigl(S^{a1}S^{bc}q^c_-\bigr)
              \bigl(\eta^{\nu\rho}\eta^{\mu\sigma}k_{3,\sigma}\bigr),
  \\[4pt]
  II  &\simeq -\frac{\delta(k)}{|z_\infty|^2}
               \bigl(S^{a1}S^{bc}q^c_-\bigr)
               \bigl(\eta^{\mu\nu}\eta^{\sigma\rho}
                     -\eta^{\mu\rho}\eta^{\sigma\nu}\bigr)k_{1,\sigma},
  \\[4pt]
  II' &\simeq -\frac{\delta(k)}{|z_\infty|^2}
               q^a_-\bigl(S^{ab}S^{c1}-S^{ac}S^{b1}\bigr)
               \bigl(\eta^{\nu\rho}\eta^{\mu\sigma}k_{3,\sigma}\bigr),
  \\[4pt]
  III &\simeq \frac{\delta(k)}{|z_\infty|^2}
              q^a_-\bigl(S^{ab}S^{c1}-S^{ac}S^{b1}\bigr)
              \bigl(\eta^{\mu\nu}\eta^{\sigma\rho}
                    -\eta^{\mu\rho}\eta^{\sigma\nu}\bigr)k_{1,\sigma} \, ,
\end{aligned}
\end{equation}
where we have defined
\begin{equation}
    S =\begin{pmatrix}
    1&0&0\\
    0&0&1\\
    0&1&0
\end{pmatrix} \, .
\end{equation}
Adding $I+II+II'+III$, including the ghost contribution $|z_\infty|^2$,
and using the transversality $A(k_i)\cdot k_{i}=0$ together with momentum
conservation, the result factorizes into a standard gauge-theory kinematic
factor times a color factor: 
\begin{equation}
  \langle\!\langle
    V^a_0(0)\,V^b_{-1}(z_\infty)\,V^c_{-1}(1)
  \rangle\!\rangle
  = \delta(k)\;t^{\mu\nu\rho}(k)\;C^{abc}(q_-),
  \label{eq:amplitude_result}
\end{equation}
where 
\begin{equation}
    \begin{aligned}
  &t^{\mu\nu\rho}(k)
  := k_{23,\sigma}\,\eta^{\rho\nu}\eta^{\mu\sigma}
    +k_{31,\sigma}\,\eta^{\mu\rho}\eta^{\nu\sigma}
    +k_{12,\sigma}\,\eta^{\mu\nu}\eta^{\rho\sigma},
    \\
  &C^{abc}(q_-)
  := \frac{1}{2}
     \bigl(q^a_-S^{ab}S^{c1}
          +q^b_-S^{bc}S^{a1}
          +q^c_-S^{ca}S^{b1}\bigr) \, 
  \label{eq:color_factor}
\end{aligned}
\end{equation}
with $k_{ab}=k_a-k_b$. The kinematic factor $t^{\mu\nu\rho}$ is the standard Yang--Mills
three-gauge-boson vertex; the color factor $C^{abc}$ encodes the flux component $f^{abc}$.  Analogous calculations for the remaining combinations of amplitudes
give 
\begin{align}
  \langle\!\langle V^{ a'}_0 V^{ b'}_{-1} V^{ c'}_{-1}\rangle\!\rangle
  &= \delta(k)\;t^{\mu\nu\rho}(k)\;C^{a'b'c'}(q'_+),
  \label{eq:amp_primes}\\[4pt]
  \langle\!\langle V^{ a'}_0 V^b_{-1} V^c_{-1}\rangle\!\rangle
  &= \delta(k)\;\bigl(k_{23,\sigma}\,\eta^{\nu\rho}\eta^{\mu\sigma}\bigr)
     \frac{q^b_+}{2}\,S^{bc}S^{a'1},
  \label{eq:amp_mixed1}\\[4pt]
  \langle\!\langle V^{a'}_0 V^{b'}_{-1} V^c_{-1}\rangle\!\rangle
  &= \delta(k)\;\bigl(k_{12,\sigma}\,\eta^{\mu\nu}\eta^{\mu\rho}\bigr)
     \frac{q'^{a'}_-}{2}\,S^{a'b'}S^{c1}.
  \label{eq:amp_mixed2}
\end{align}
The non-standard kinematic factor in the mixed amplitudes
\eqref{eq:amp_mixed1} and \eqref{eq:amp_mixed2} (a single $k_{ab}$ tensor
rather than $t^{\mu\nu\rho}$) is the signature of a \emph{non-compact}
algebra.

Matching the amplitudes to those of a gauge field theory with $O(3,3)$
symmetry, one reads off the structure constants in the effective theory: 
\begin{equation}
  \begin{aligned}
    f_{abc}   &= C^{abc}(q_-)
               = \tfrac{1}{2}\bigl(q^b_-S^{bc}S^{a1}
                                  +q^c_-S^{ca}S^{b1}
                                  +q^a_-S^{ab}S^{c1}\bigr),\\[4pt]
    f_{a'b'c'}&= -C^{a'b'c'}(q'_+)= -\tfrac{1}{2}\bigl(q'^b_+ S^{b'c'}S^{a'1}
                                  +q'^c_+ S^{c'a'}S^{b'1}
                                  +q'^a_+ S^{a'b'}S^{c'1}\bigr),\\[4pt]
    f_{a'bc}  &= \tfrac{1}{2}q^b_+S^{bc}S^{a'1}, \\[4pt]
    f_{ab'c}  &= \tfrac{1}{2}q^c_+S^{ca}S^{b'1}, \\[4pt]
    f_{abc'}  &= \tfrac{1}{2}q^a_+S^{ab}S^{c'1},\\[4pt]
    f_{ab'c'} &= -\tfrac{1}{2}q'^{b'}_-S^{b'c'}S^{a1},\\[4pt]
    f_{a'bc'} &= -\tfrac{1}{2}q'^{c'}_-S^{c'a'}S^{b1},\\[4pt]
    f_{a'b'c} &= -\tfrac{1}{2}q'^{a'}_-S^{a'b'}S^{c1}.
  \end{aligned}
  \label{eq:structure_constants}
\end{equation}
Moving to the up/down basis\footnote{By abuse of notation we use the same notation $a$ for left-moving components and up/down indices.} 
\begin{equation}
  A^a_\mu = A_{\mu,a}+A'_{\mu,a'},
  \qquad
  A_{\mu a} = A_{\mu,a}-A'_{\mu,a'},
  \label{eq:updown_basis}
\end{equation}
we find the components of the generalized flux to be
\begin{equation}
\begin{aligned}
        &H_{abc}=\frac{1}{2}\left(q_{-}-q'_{+}+q_{+}-q'_{-}\right)^bS^{bc}S^{a1}\pm \text{perm.},
        \\
        &\omega_{ab}^{\ \ c}=\frac{1}{2}\left(q_{-}+q'_{+}+q_{+}+q'_{-}\right)^bS^{bc}S^{a1}+\frac{1}{2}\left(q_{-}+q'_{+}+q_{+}+q'_{-}\right)^cS^{ac}S^{b1}\\
        &\qquad \ \ +\frac{1}{2}\left(q_{-}+q'_{+}-q_{+}-q'_{-}\right)^aS^{ab}S^{c1},\\
        &Q_{a}^{\ bc}=\frac{1}{2}\left(q_{-}-q'_{+}+q_{+}-q'_{-}\right)^bS^{bc}S^{a1}+\frac{1}{2}\left(q_{-}-q'_{+}-q_{+}+q'_{-}\right)^cS^{ac}S^{b1}\\
        &\qquad \ \ +\frac{1}{2}\left(q_{-}-q'_{+}-q_{+}+q'_{-}\right)^aS^{ab}S^{c1},
        \\
        &R^{abc}=\frac{1}{2}\left(q_{-}+q'_{+}-q_{+}-q'_{-}\right)^bS^{bc}S^{a1}\pm\text{perm.}\ .
\end{aligned}
\end{equation}
Using the explicit expressions of $q$ and $q'$ in terms of the orbifold parameters \eqref{eq:qa_pm_explicit} we get
\begin{equation}
\begin{aligned}
        &\left(H_{123},Q_1^{\ 23},Q_2^{\ 31},Q_3^{\ 12}\right)=
   - \left(\frac{2\lambda_{-}}{R},\frac{2\lambda_{-}}{R},0,0\right),\\
     &\left(R^{123},\omega_{23}^{\ \ 1},\omega_{31}^{\ \ 2},\omega_{12}^{\ \ 3}\right)=
     \left(0,0,\frac{2\lambda_{+}}{R},\frac{2\lambda_{+}}{R}\right),
    \label{dict_dim}
\end{aligned}
\end{equation}
where 
\begin{equation}
\lambda_\pm=\lambda_L\pm\lambda_R. 
\end{equation}
The $R$ dependency can be removed by redefining\footnote{The effective radius in the low-energy theory for the orbifolded space is $\frac{R}{2N}$ and not just $R$.} $A_{\mu}^1=\frac{R}{2N}\, \tilde{A}_{\mu}^1$ and $A_{\mu,1}=\frac{\tilde{A}_{\mu,1}}{R/2N}$, leading to 
\begin{equation}
\begin{aligned}
        &\left(H_{123},Q_1^{\ 23},Q_2^{\ 31},Q_3^{\ 12}\right)=
   - \left(\frac{\lambda_{-}}{N},\frac{\lambda_{-}}{N}, 0,0\right),\\
     &\left(R^{123},\omega_{23}^{\ \ 1},\omega_{31}^{\ \ 2},\omega_{12}^{\ \ 3}\right)=
     \left(0,0,\frac{\lambda_{+}}{N},\frac{\lambda_{+}}{N}\right).
    \label{dict}
\end{aligned}
\end{equation}
Note that the flux is quantized in every case,
since the volume of $T^3/\mathbb Z_N$ is $1/N$ times that of $T^3$. 

These expressions
provide the dictionary between the orbifold twist eigenvalues and the
generalized fluxes, and are the key result used to match orbifold models
to the gauged supergravity entries in table~\ref{tab:spectra}; with the identifications
\begin{equation}
\alpha=-\frac{\lambda_-}{N}, \quad \,\beta=\frac{\lambda_+}{N}.  
\label{alphabetalambda}
\end{equation}
 We stress these are the only possible fluxes arising from a modular invariant, freely acting abelian orbifold of $T^3$. 
\\

A further check is the matching of the masses of scalar fields. Their vertex operators in the $(-1)$-picture are:
\begin{equation}
    V^{ab}_{-1}=\psi^a \, \bar\psi^b \, e^{ik\cdot X} e^{i Q^{ab}_{+}y_L+iQ^{ab}_{-}y_R},
\end{equation}
with $Q^{ab}_{\pm}=\frac{m^{ab}}{R}\pm R\,n^{ab}$, and $a,b=1,2,3$.
The invariance under the orbifold action together with the on-shell condition result in\footnote{We use the same notation $m_{ab}$ for the mass of the scalars $\phi_{ab}$ and their momentum number. The distinction should be clear from the context.} 
\begin{equation}
    Q^{ab}_{\pm}\overset{!}{=}\frac{2(\lambda_L^a+\lambda_R^b)}{R}, \qquad -k^2=\tilde{{m}}_{ab}^2\overset{!}{=}\frac{4(\lambda_L^a+\lambda_R^b)^2}{R^2}.
\end{equation}
Notice that $\tilde{m}_{ab}^2$ are the masses of the scalar fields $\tilde{\phi}_{ab}$ in a dimensionful basis. To move to a dimensionless basis we define the fields $\phi_{ab}=\frac{\tilde{\phi}_{ab}}{R/2N}$, using the effective radius, so that:
\begin{equation}
    m_{ab}^2=\left(\frac{\lambda_L^a+\lambda_R^b}{N}\right)^2.
\end{equation}
From the explicit values of the twists we can read off the scalar masses:
\begin{equation}
    \begin{aligned}
    &m_{11}^2=0,\\
    &m_{12}^2=m_{13}^2=\left(\frac{\lambda_R}{N}\right)^2=\frac{\left(\alpha+\beta\right)^2}{4},\\
    &m_{21}^2=m_{31}^2=\left(\frac{\lambda_L}{N}\right)^2=\frac{\left(\alpha-\beta\right)^2}{4},\\
    &m_{23}^2=m_{32}^2=\left(\frac{\lambda_L+
    \lambda_R}{N}\right)^2=\beta^2,\\
    &m_{22}^2=m_{33}^2=\left(\frac{\lambda_L-\lambda_R}{N}\right)^2=\alpha^2.
\end{aligned}
\label{masses-orbifold}
\end{equation}
As expected, the circle volume modulus is massless. All the scalars with one leg along the circle, if not massless by the condition $\alpha=\pm \beta$, have the same mass as the ones eaten by the gauge bosons to get mass in the effective theory. The remaining massive scalars are the actual scalars retained in the effective theory.

\section{The landscape and the swampland}
\label{sec:landscape_swampland}

We are now ready to inspect the various entries of table \ref{tab:spectra}, and consider the actions they generate in order to see if they are realizable as freely acting orbifolds, whose gaugings are given in \eqref{dict}. This is possible when the gauging admits a vacuum with $V=0$, so we restrict to those.

\paragraph{8,7:}
This gauging can be realized as an orbifold only for the following gauging parameters \footnote{These orbifold actions are T-dual to the ones given in table \ref{tab:rotsym} which have the same sign for $\lambda_L$ and $\lambda_R$.}:
\begin{equation}
\alpha=2\frac{\,\lambda_{R}}{N}=-2\frac{\,\lambda_{L}}{N}=\{1,\frac{2}{3},\frac{4}{3},\frac{1}{2},\frac{1}{3}\}, \quad \quad \beta=0.
\end{equation}
The masses of the scalars for this gaugings also coincide with those computed from the orbifold, given in \eqref{masses-orbifold}.
There are thus five symmetric $\mathbb{Z}_{2N}$ orbifolds that match this gauged supergravity configuration. 
Level matching, and thus modular invariance, whose condition is given in \eqref{eq:ModInv}, is  trivially satisfied in this case as the twists are symmetric and there is no invariant lattice $I$, implying therefore $v^*_{L(R)}=0$.

This gauged supergravity can also be obtained from a reduction on an ISO(2) manifold, as explained in section \ref{sec:GaugingsST}. Such manifold is compact  only for $\alpha=1/N$, similarly to the orbifold case. Note however that for an $ISO(2)$ manifold $N$ can be any natural number, while here we see, not surprisingly, that only for a very restricted set of $\alpha$'s the vacuum of this Scherk-Schwarz compactification can be realized by an orbifold. 

\paragraph{6:}
This configuration has the form \eqref{dict}, where the gaugings parameters are relate to the orbifold actions by 
\begin{equation}
    \alpha=-\frac{\,(\lambda_{L}-\lambda_{R})}{N}, \quad \beta=\frac{(\lambda_{L}+\lambda_{R})}{N}.
\end{equation}
While this gauging has a Minkowski vacuum for any $\alpha$ and $\beta$, only very few choices, namely those corresponding to the $\lambda$'s given in table \ref{tab:rotsym}, can be realized. Those are given by the following:
\begin{itemize}
    \item $\alpha=\beta$: 
\begin{equation}
    \quad\lambda_L=0, \quad \beta=\alpha=\,\frac{\lambda_R}{N}=\{\frac{1}{2},\frac{1}{3},\frac{2}{3}\}.
    \label{eq:aupb}
\end{equation}
These correspond to a $\mathbb{Z}_4$ and two $\mathbb{Z}_6$ asymmetric orbifolds.  Level matching is trivially satisfied since we doubled the order of the orbifold due to the fermions. Note that for $\lambda/N=2/3$ one could equally do an order 3 shift. 
   
    \item $\alpha=-\beta$. This case is slightly more subtle. If one blindly follows \eqref{dict}, one finds $\lambda_R=0$ and $\alpha=-\frac{\lambda_L}{N}$. A purely left action (with the appropriate values of $\lambda$) generates however a supersymmetric spectrum, actually the same as that of the previous situation. However, as indicated in table \ref{tab:spectra}, this gauged supergravity vacuum is non supersymmetric, and therefore this cannot be its stringy realization. The solution to this issue is to add to the orbifold  a $(-1)^{F_R}$ action, with $F_{R}$ the spacetime fermion number in the right-moving sector. This is precisely a chiral $2\pi$ rotation on the bosons which acts on the right-moving fermions as $(-1)^{F_R}$. To realize this "trivial" rotation we should set $\lambda_R=N$. 
Therefore solving \eqref{eq:inv} for the lowest possible $m^a$ and $n^a$, that is both zero, one finds the flux configuration 
\begin{equation}
    \lambda_R=N, \quad \beta=-\alpha=\,\frac{\lambda_L}{N}=\{\frac{1}{2},\frac{1}{3},\frac{2}{3}\}.
\end{equation}
Also in this case the level matching is trivially satisfied.
    \item For $\beta\neq \alpha$, the only possibility realisable is the quasicrystalline $\mathbb{Z}_{12}$ action, for which
\begin{equation}
     \alpha=-\frac{(1-5)}{12}=\frac{1}{3}, \quad \beta=\frac{(1+5)}{12}=\frac{1}{2}.
\end{equation}
Also in this case the orbifold order is doubled to $\mathbb{Z}_{24}$. Level matching can be easily checked.
\end{itemize}

\noindent For all of these cases, the scalar masses of the gauged supergravity vacuum match those or the orbifold, given in \eqref{masses-orbifold}.

\paragraph{3,1:}
Despite the fact that these gauged supergravities have a Minkowski vacuum for $\alpha=-\beta$ and $\alpha=\beta$, they cannot be realized as orbifolds since their fluxes are not of the form  \eqref{dict}. 
It is possible that extending the construction of section \ref{sec:vacuaCFT} to non-abelian point groups could give the stringy realisation of these two vacua.

\subsection{Explicit realisations of the flux configurations}

In this subsection we present the explicit string theory realization of the vacua 8, 7 and 6  in table \ref{tab:spectra}, according to the results of the previous section. 
\label{sec:explicitOrb}
\subsection*{Symmetric orbifold (8 and 7):}

The partition function\footnote{For the sake of simplicity we omitted the non compact bosons contribution $\left(\sqrt{\tau_2}\ \bar{\eta}\eta\right)^5.$} of the type II theories compactified on $T^2\times S^1$. It can be written in terms of Jacobi eta and theta functions\footnote{See appendix \ref{App:conventions} for the definition of the modular functions.} as follows:
\begin{equation}
	\mathcal{Z}\left[\genfrac{}{}{0pt}{}{0}{0}\right]= \left| 
	\frac{1}{2} \sum_{a,b}	\eta_{ab} \ \frac{\theta^3\left[\genfrac{}{}{0pt}{}{\nicefrac{a}{2}}{\nicefrac{b}{2}}\right]}{\eta^3}	
	\frac{\theta\left[\genfrac{}{}{0pt}{}{\nicefrac{a}{2}}{\nicefrac{b}{2}}\right]}{\eta}	
	\right|^2 	\Gamma\left[\genfrac{}{}{0pt}{}{0}{0}\right]\Lambda\left[\genfrac{}{}{0pt}{}{0}{0}\right],
	\label{z00}
\end{equation}

\noindent
where we have already separated the spacetime theta functions from the internal one, and the GSO projection is $\eta_{ab}=\left(-1\right)^{a+b+ab}$. $\Gamma\left[\genfrac{}{}{0pt}{}{0}{0}\right]$ and $\Lambda\left[\genfrac{}{}{0pt}{}{0}{0}\right]$ are the Narain lattices of the torus and the circle respectively:

\begin{equation}
		\Gamma\left[\genfrac{}{}{0pt}{}{0}{0}\right]=\sum_{\vec{m},\vec{n}}\frac{\Lambda^{^{(2)}}_{\vec{m},\vec{n}}}{\left(\bar{\eta} \eta\right)^2}, \qquad \qquad
			\Lambda\left[\genfrac{}{}{0pt}{}{0}{0}\right]=\sum_{m,n}\frac{\Lambda^{^{(1)}}_{m,n}}{\left(\bar{\eta} \eta\right)},
\end{equation}

\noindent
and are defined explicitely in appendix \ref{App:conventions}.
The orbifold action  on the bosonic sector is a $\mathbb{Z}_N$ crystallographic action  ($N=2,3,4,6$)\footnote{To avoid cumbersome notation we are omitting the possibility of a $\frac{4\pi}{3}$ twist, which is actually the only one that realizes also $Z_3$ action also on the fermions, as the resulting model is the same as the other $N\neq2$.} on $T^2$ accompanied by a shift of order $2N$ -on the world-sheet fermions the orbifold is a $Z_{2N}$ action, as we will see in a moment. Indeed, as already saw in the general construction at the beginning of the section, in order to preserve the world-sheet supercurrent an analogous action on the relative world-sheet fermions is mandatory. If we call $\psi$ the complex fermion superpartner of the $T^2$ bosons, one needs:
\begin{equation}
	\psi\longrightarrow e^{2 i\pi g/N} \ \psi \qquad g=0,\cdots N-1.
	\label{actpsi}
\end{equation}
It is now possible to bosonize the world-sheet fermions and build the spin field as:
\begin{equation}
	S = e^{\frac{i}{2}\sum_{i=1}^4\epsilon_i H_i}, \qquad \epsilon_i=\pm 1, \qquad i=1,... \;,4,
\end{equation}
where $H_i$ are the bosonized fermions. Calling $H_4$ the bosonization of the world-sheet fermions along $T^2$, the action \eqref{actpsi} translates in $H_4\longrightarrow H_4+\frac{2\pi g}{N}$, and thus
\begin{equation}
	S\longrightarrow e^{i\pi \epsilon_4 g/N} \ S.
\end{equation}
Recalling that the spin field is the generator of the spacetime supersymmetry, we see that all the supercharges are projected-out by the orbifold action. Moreover, for $g=N$ one has:
\begin{equation}
	S\longrightarrow \left(-1\right)^{\epsilon_4} S,
\end{equation}
which is the action of the spacetime fermion number, and in particular it is not the identity.
So the lift of the bosonic $\mathbb{Z}_N$ action to the fermions produces effectively a $\mathbb{Z}_{2N}$ orbifold. The partition function on the first twisted sector is given by
\begin{equation}
	\mathcal{Z}\left[\genfrac{}{}{0pt}{}{0}{g}\right]=  \left| 
	\frac{1}{2} \sum_{a,b}	\eta_{ab} \ \frac{\theta^3\left[\genfrac{}{}{0pt}{}{\nicefrac{a}{2}}{\nicefrac{b}{2}}\right]}{\eta^3}	
	\frac{\theta\left[\genfrac{}{}{0pt}{}{\nicefrac{a}{2}}{\nicefrac{b}{2}+\nicefrac{g}{N}}\right]}{\eta}	
	\right|^2 	\Gamma\left[\genfrac{}{}{0pt}{}{0}{g}\right]\Lambda\left[\genfrac{}{}{0pt}{}{0}{g}\right].
\end{equation}
The full partition function can be straightforwardly obtained by applying the modular transformations on the above expression, and then summing over all sectors:
\begin{equation}
	\begin{split}
		\mathcal{Z}&=\frac{1}{2N}\sum_{h,g} \mathcal{Z}\left[\genfrac{}{}{0pt}{}{h}{g}\right]\\
		&= \frac{1}{\left(\sqrt{\tau_2}\bar{\eta}\eta\right)^5}\left| 
		\frac{1}{2} \sum_{a,b}	\eta_{ab} \ \frac{\theta^3\left[\genfrac{}{}{0pt}{}{\nicefrac{a}{2}}{\nicefrac{b}{2}}\right]}{\eta^3}	
	\frac{\theta\left[\genfrac{}{}{0pt}{}{\nicefrac{a}{2}+\nicefrac{h}{N}}{\nicefrac{b}{2}+\nicefrac{g}{N}}\right]}{\eta}	\ e^{-i\pi \frac{h}{N}(b+\frac{g}{N})}
		\right|^2 	\Gamma\left[\genfrac{}{}{0pt}{}{h}{g}\right]\Lambda\left[\genfrac{}{}{0pt}{}{h}{g}\right],
	\end{split}
		\label{zhg} 
\end{equation}

\noindent
where we have reintroduced the contribution of the spacetime bosons. The full expressions for $\Gamma$ and $\Lambda$ are:
\begin{equation}
	\begin{split}
		\Lambda\left[\genfrac{}{}{0pt}{}{h}{g}\right]&=\sum_{m,n} e^{i \pi gm/N}\ \frac{\Lambda^{^{(1)}}_{m,n+\frac{h}{2N}}}{\bar{\eta} \eta},
		\\
		\Gamma\left[\genfrac{}{}{0pt}{}{h}{g}\right]&=
		\begin{cases}
			\sum_{\vec{m},\vec{n}}\frac{\Lambda^{^{(2)}}_{\vec{m},\vec{n}}(T_*,U_*)}{\left(\bar{\eta} \eta\right)^2}, \quad\qquad h=g=0,
			\\
			\left|	\frac{2 \sin\left(\frac{\pi}{N} (h,g)\right) \eta}{\theta\left[
            \genfrac{}{}{0pt}{}{\nicefrac{1}{2}-\nicefrac{h}{N}}{\nicefrac{1}{2}-\nicefrac{g}{N}}
            \right]}	\right|^2, \qquad \quad \ \text{otherwise}\, ,
		\end{cases}
	\end{split}
\end{equation}
where $(h,g)$ stands for the greatest common divisor and the values $T_*$ and $U_*$  are the ones compatible with the specific action, and are given in table \ref{tab:rotsym}.

Due to the shift on the circle, the twisted states are  massive for sufficiently large radius, which is the regime of validity of the effective field theory. Therefore, in the search for massless states we can focus on the untwisted sector. 
Using the relations in appendix \ref{App:conventions}, one can write the spacetime theta functions in terms of the $SO(6)$ level one affine character, which is helpful to read off the spectrum. Keeping only the possible massless contributions, the q-expanded partition function is:
\begin{equation}
	\mathcal{Z}\approx\frac{1}{\left(\sqrt{\tau_2}\bar{\eta}\eta\right)^5} \frac{1}{\left(\bar{\eta}\eta\right)^4}
	\left[
	|V_6|^2 + n_s\left(N\right) |O_6|^2 \left(\bar{q}q\right)^\frac{1}{2}+ \left(|S_6|^2+|C_6|^2\right)\left(\bar{q}q\right)^\frac{1}{8}
	\right],
	\label{spectr}
\end{equation} 
where the number of massless scalar depends on the model: $n_s\left(2\right)=4$, $n_s\left(N\neq2\right)=2$.
From this one can easily read the 8D content of the spectrum and reduce it to 7D. In particular, one has:
\begin{equation}
	\begin{split}
			&|V_6|^2\longrightarrow g_{MN}, B_{MN}, \Phi,
			\\
			&|O_6|^2\longrightarrow\phi,
			\\
			&|S_6|^2+|C_6|^2\longrightarrow 2\left(C_0+C_{MN}\right),
	\end{split}
\end{equation}
where $M,N$ are 8D indices. One can see that the 8D massless spectrum consist of a metric field $g_{MN}$, three 2-forms $B_{MN},\  2C_{MN}$, and $(n_s+3)$ scalars: $\Phi, \ n_s \phi,\  2C_0$. The reduction to 7D yields a metric field, three 2-forms, four vectors and ($n_s+4$) scalars. 

The further truncation of massless spectrum to the NSNS sector results in the following field content:
\begin{equation}
		\begin{split}
			g_{\mu\nu},B_{\mu\nu},\Phi, 2A_\mu,(n_s+1)\phi,
		\end{split}
\end{equation}

\noindent
where $\mu,\nu$ are 7D indices. Considering the orbifolds for $N\neq2$, one can easily verify that the massless spectrum agrees with the ones of the entries 7 and 8 in Table \ref{tab:spectra}. As shown, also the masses of the lightest boson match those in the Table. 

\subsection*{Asymmetric and quasicrystalline orbifolds (6):} 
The orbifold realization of the various configurations of fluxes in the vacuum 6 in table \ref{tab:spectra} is given by

\paragraph*{$\alpha=0 \vee \beta=0$:}This configuration of 6 just boils down to 7,8, symmetric orbifolds.

\paragraph*{$\beta=\alpha$:}
These are truly asymmetric orbifolds. The only possibilities are given in Table \ref{tab:rotsym}, which lead to \eqref{eq:aupb}. These can be T-dualized to purely left actions $\lambda_L/N=\{\frac{1}{2},\frac{1}{3},\frac{2}{3}\}$. Since there is no action on the right-moving sector, in particular on the right-moving fermions, the vacuum is $\mathcal{N}=1$ supersymmetric in 7D.
As an example\footnote{See appendix \ref{App:other} for the other two models.}, we show the $\lambda=\tfrac12$ case, which fixes the $T^2$ lattice to be at the $SO(4)_L\times SO(4)_R$ point in moduli space. The metric on $T^2$, which is 1/2 the Cartan matrix of $SO(4)$, is the identity matrix, and the internal B-field, 1/2 of its anti-symmetrized version, is zero. 
At this very special point in moduli space (the fermionic point) we can fermionize the $T^2$ bosonic degrees of freedom, and write their contribution to the partition function in terms of theta functions:

\begin{equation}
	\Gamma\left[\genfrac{}{}{0pt}{}{0}{0}\right] =\frac{1}{2}\sum_{\delta,\gamma}\left| \frac{\theta\left[\genfrac{}{}{0pt}{}{\nicefrac{\gamma}{2}}{\nicefrac{\delta}{2}}\right]}{\eta} \right|^4.
    \label{eq:so4lat}
\end{equation}

\noindent
The advantage of the feminization is that the rotation on the compact left-moving bosons becomes a shift on the lower characteristic of one of the left moving theta in \eqref{eq:so4lat}.
Of course also an analogous action on the worldsheet fermions is needed, as in the previous symmetric case. Notice that in this asymmetric case the $\mathbb{Z}_2$ action already at the level of the bosons is realized as a  $\mathbb{Z}_4$ action. 
The projected partition function is the following:

\begin{equation}
	\begin{split}
			\mathcal{Z}\left[\genfrac{}{}{0pt}{}{0}{1}\right]=&  \
		\frac{1}{2} \sum_{\bar{a},\bar{b}}	\eta_{\bar{a}\bar{b}} \ \frac{\bar{\theta}^4\left[\genfrac{}{}{0pt}{}{\nicefrac{\bar{a}}{2}}{\nicefrac{\bar{b}}{2}}\right]}{\bar{\eta}^4}	
		\ \frac{1}{2} \sum_{a,b}	\eta_{ab} \ \frac{\theta^3\left[\genfrac{}{}{0pt}{}{\nicefrac{a}{2}}{\nicefrac{b}{2}}\right]}{\eta^3}	
		\frac{\theta\left[\genfrac{}{}{0pt}{}{\nicefrac{a}{2}}{\nicefrac{b}{2}+\nicefrac{1}{2}}\right]}{\eta}	
		\\
		&\ \, \frac{1}{2}\sum_{\delta,\gamma}\frac{\bar{\theta}^2\left[\genfrac{}{}{0pt}{}{\nicefrac{\delta}{2}}{\nicefrac{\gamma}{2}} \right]}{\bar{\eta}^2} \frac{\theta\left[\genfrac{}{}{0pt}{}{\nicefrac{\delta}{2}}{\nicefrac{\gamma}{2}} \right]\theta\left[\genfrac{}{}{0pt}{}{\nicefrac{\delta}{2}}{\nicefrac{\gamma}{2}+\nicefrac{1}{2}} \right]}{\eta^2}\Lambda\left[\genfrac{}{}{0pt}{}{0}{g} \right].
	\end{split}
	\label{fasymz2}
\end{equation}

\noindent
The full partition function is then easily obtained by modular completing this piece. The q-expansion up to massless states finally reads:

\begin{equation}
	\mathcal{Z}\approx\frac{1}{\left(\sqrt{\tau_2}\bar{\eta}\eta\right)^5} \frac{1}{\left(\bar{\eta}\eta\right)^4}
	\left[
	|V_6|^2 + 2 \bar{V}_6 O_6 \ q^\frac{1}{2}- \left(\bar{S}_6+\bar{C}_6\right)V_6 \ \bar{q}^\frac{1}{8}
	\right].
\end{equation}

\noindent
In this case the 7 dimensional spectrum consists of a gravity and a vector multiplet. The former contains the metric, the B-field, the dilaton, three vectors, a gravitino and a spin 1/2 fermion, and the latter a vector, three scalars and a fermion. The above spectrum is exactly the massless spectrum of 6 for $\beta=\alpha$.

\subsubsection*{$\beta=-\alpha$:}
As discussed in the previous section, this case is obtained from the previous orbifold just by "adding" a $(-1)^{F_R}$ to the previous $Z_2$ action on the world-sheet fermions.

We give again the explicit construction of  the $\mathbb{Z}_2$ action (effectively $\mathbb{Z}_4$ as before). The untwisted projected partition function is given by
\begin{equation}
	\begin{split}
			\mathcal{Z}\left[\genfrac{}{}{0pt}{}{0}{1}\right]=&  \
		\frac{1}{2} \sum_{\bar{a},\bar{b}}	\eta_{\bar{a}\bar{b}}(-1)^{\bar{a}} \ \frac{\bar{\theta}^4\left[\genfrac{}{}{0pt}{}{\nicefrac{\bar{a}}{2}}{\nicefrac{\bar{b}}{2}}\right]}{\bar{\eta}^4}	
		\ \frac{1}{2} \sum_{a,b}	\eta_{ab} \ \frac{\theta^3\left[\genfrac{}{}{0pt}{}{\nicefrac{a}{2}}{\nicefrac{b}{2}}\right]}{\eta^3}	
		\frac{\theta\left[\genfrac{}{}{0pt}{}{\nicefrac{a}{2}}{\nicefrac{b}{2}+\nicefrac{1}{2}}\right]}{\eta}	
		\\
		&\ \, \frac{1}{2}\sum_{\delta,\gamma}\frac{\bar{\theta}^2\left[\genfrac{}{}{0pt}{}{\nicefrac{\delta}{2}}{\nicefrac{\gamma}{2}} \right]}{\bar{\eta}^2} \frac{\theta\left[\genfrac{}{}{0pt}{}{\nicefrac{\delta}{2}}{\nicefrac{\gamma}{2}} \right]\theta\left[\genfrac{}{}{0pt}{}{\nicefrac{\delta}{2}}{\nicefrac{\gamma}{2}+\nicefrac{1}{2}} \right]}{\eta^2}\Lambda\left[\genfrac{}{}{0pt}{}{0}{g} \right].
	\end{split}
	\label{sasym}
\end{equation}
Also in this case the modular completion is easily obtainable. The q-expansion yields:

\begin{equation}
	\mathcal{Z}\approx\frac{1}{\left(\sqrt{\tau_2}\bar{\eta}\eta\right)^5} \frac{1}{\left(\bar{\eta}\eta\right)^4}
	\left[
	|V_6|^2 + 2 \bar{V}_6 O_6 \ q^\frac{1}{2}- \left(\bar{S}_6+\bar{C}_6\right)O_6 \ q^{\frac{1}{2}} \bar{q}^\frac{1}{8}
	\right].
\end{equation}

\noindent
The massless spectrum in this case is composed by the metric, the B-field, the dilaton, four vector, three scalars, and two spin 1/2 fermions, which is non-supersymmetric and matches the massless spectrum of 6 for $\beta=-\alpha$.

\subsubsection*{$\beta\neq \pm \alpha$:}
As explained, the only possibility (up to T-duality) is $\beta=\frac{1}{2}, \ \alpha=\frac{1}{3}$. This corresponds to a quasi-crystallographic $\mathbb{Z}_{12}$ action that, as in the previous cases, is lifted to the fermions as a $\mathbb{Z}_{24}$. This action fixes the moduli to  $T_*=\frac{1}{2}+i \frac{\sqrt{3}}{2}$ and $U_*=i$. 
The modular invariant partition function reads:
\begin{equation}
	\begin{split}
		\mathcal{Z}&=\frac{1}{24}\sum_{h,g} \mathcal{Z}\left[\genfrac{}{}{0pt}{}{h}{g}\right]\\
		&= \frac{1}{\left(\sqrt{\tau_2}\bar{\eta}\eta\right)^5}
    \
		\frac{1}{2} \sum_{a,b}	\eta_{ab} \ \frac{\theta^3\left[\genfrac{}{}{0pt}{}{\nicefrac{a}{2}}{\nicefrac{b}{2}}\right]}{\eta^3}	
	\frac{\theta\left[\genfrac{}{}{0pt}{}{\nicefrac{a}{2}+\nicefrac{h}{12}}{\nicefrac{b}{2}+\nicefrac{g}{12}}\right]}{\eta}	\ e^{-i\pi \frac{h}{12}(b+\frac{g}{12})}
    \\ 
    &  \quad  
    \frac{1}{2} \sum_{\bar{a},\bar{b}}	\eta_{\bar{a}\bar{b}} \ \frac{\bar{\theta}^3\left[\genfrac{}{}{0pt}{}{\nicefrac{\bar{a}}{2}}{\nicefrac{\bar{b}}{2}}\right]}{\bar{\eta}^3}	
	\frac{\theta\left[\genfrac{}{}{0pt}{}{\nicefrac{\bar{a}}{2}+\nicefrac{5h}{12}}{\nicefrac{\bar{b}}{2}+\nicefrac{5g}{12}}
   \right]}{\bar{\eta}}	\ e^{i\pi \frac{5h}{12}(\bar{b}+\frac{5g}{12})} \;
    \Gamma\left[\genfrac{}{}{0pt}{}{h}{g}\right] \;  \Lambda\left[\genfrac{}{}{0pt}{}{h}{g}\right],
	\end{split}
		\label{zhgas}
\end{equation}
where:
\begin{equation}
	\begin{split}
		\Lambda\left[\genfrac{}{}{0pt}{}{h}{g}\right]&=\sum_{m,n} e^{i \pi gm/12}\ \frac{\Lambda^{^{(1)}}_{m,n+\frac{h}{24}}}{\left(\bar{\eta} \eta\right)},&
		\\
		\Gamma\left[\genfrac{}{}{0pt}{}{h}{g}\right]&=
		\begin{cases}
			\sum_{\vec{m},\vec{n}}\frac{\Lambda^{^{(2)}}_{\vec{m},\vec{n}}(T_*,U_*)}{\left(\bar{\eta} \eta\right)^2}, \quad \qquad \qquad \qquad \quad \ \quad h=g=0,
			\\ \\
			\frac{2 \sin\left(\frac{\pi}{12} (h,g)\right) \eta}{\theta\left[\genfrac{}{}{0pt}{}{\nicefrac{1}{2}-\nicefrac{h}{12}}{\nicefrac{1}{2}-\nicefrac{g}{12}}\right]}
            \
            \frac{2 \sin\left(\frac{5\pi}{12} (h,g) \right) \bar{\eta}}{\bar{\theta}\left[\genfrac{}{}{0pt}{}{\nicefrac{1}{2}-\nicefrac{5h}{12}}{\nicefrac{1}{2}-\nicefrac{5g}{12}}\right]} e^{i \pi \frac{h}{3}(1-\frac{g}{2})}
            , \qquad \text{otherwise},
		\end{cases}
	\end{split}
\end{equation}

\noindent
where $(h,g)$ is the greatest common divisor of h and g. The q-expantion up to massless states reads:

\begin{equation}
	\mathcal{Z}\approx\frac{1}{\left(\sqrt{\tau_2}\bar{\eta}\eta\right)^5} \frac{1}{\left(\bar{\eta}\eta\right)^4}
	|V_6|^2.
\end{equation}
\noindent
The massless spectrum then consist of the metric, the B-field, the dilaton, two vectors and one scalar, which matches the spectrum of the general case of 6. 	
	
	\section{Conclusions}

In this paper we used the complete classification of half-maximal gauged supergravities
in $D=7$~\cite{dibitetto_duality_2012,dibitetto_all_2015} to address a concrete instance of
the landscape/swampland question: which of these gaugings are in the landscape and thus admit an ultraviolet completion, and which do not. Most entries in Table \ref{tab:gaugings} involve non-geometric fluxes that
cannot be dualized away to a frame in which the compactification is purely geometric, so it
is essential to determine which of these non-geometric configurations has nonetheless a
string theory realization.

We explored the corner of the non-geometric landscape of freely acting asymmetric
orbifolds of $T^3$. By computing tree-level three-point amplitudes of the vector fields in the
orbifolded theory, we built an explicit dictionary between the orbifold twist eigenvalues and
the $O(3,3)$ generalized fluxes, summarized in~\eqref{dict}. This dictionary turns out to
be remarkably rigid: the only fluxes that can be turned on by a modular-invariant, freely
acting $\mathbb Z_N$ orbifold of $T^3$ take the simple form $H,Q,\omega\propto 1/N$, with
$N\in\{2,3,4,6,12\}$, and $R=0$. The first four values of $N$ correspond to the crystallographic point groups of
the $T^2$ lattice, while $N=12$ is realized only by a quasicrystalline action. Note that the flux is quantized in every case,
since the volume of $T^3/\mathbb Z_N$ is $1/N$ times that of $T^3$.

We showed that this class of orbifolds can never generate
\emph{intrinsic $R$-flux}, by which we mean $R$-flux that survives in every T-duality frame
rather than being dualizable away together with the other fluxes. As a result, only a single
entry of Table~1, namely gauging 6, admits a
complete string realization through this construction, and only for $\beta=0, \alpha=\pm\beta$, $\alpha=\frac23 \beta$, corresponding respectively to symmetric, asymmetric and quasicrystalline actions.  Futhermore, there are only a handfulof allowed values for  $\alpha$ and $\beta$. Any other gauging that carries intrinsically
non-geometric flux and has a vacuum with $V=0$ (as expected for freely acting orbifods since they are flat), but is not realized by the dictionary \eqref{dict}, is therefore a strong candidate for the gauged-supergravity swampland.

It is worth noting that combining two independent orbifold actions, one
involving a momentum shift and another one a winding shift, can in principle generate $R$-flux
directly \cite{condeescu_gauged_2013}. On $T^3$, however, such combined actions, whenever they generate, in principle, intrinsic $R$-flux, are not
modular invariant. It remains an open question whether this is a
genuine obstruction or merely an artifact of the low dimensionality of $T^3$: on
higher-dimensional tori, additional independent twists and shift directions are possible, and it is
conceivable that a modular-invariant asymmetric orbifold realizing intrinsic $R$-flux could be
constructed there. We leave this direction for future work.

There are two natural further extensions of this analysis. First, our discussion has
deliberately excluded gaugings involving the $\mathbb R^+$ factor of the duality group, whose
embedding tensor is in the fundamental of $O(3,3)$. 
Extending the landscape/swampland classification to include these would complete the
picture for the full $\mathbb R^+\times O(3,3)$ duality group. Second, we have restricted
throughout to abelian orbifold actions. Non-abelian asymmetric orbifolds are a
significantly richer and largely unexplored class of constructions, and have recently been the
subject of renewed interest, both for vanishing one-loop vacuum
energy~\cite{kakushadze_asymmetric_1996,larotonda_asymmetric_2026,fraiman_nonabelian_2026} and,
earlier, in connection with $T$-fold and non-geometric flux backgrounds via chiral orbifold
twists~\cite{satoh_nonsusy_2015}. To our knowledge, however, the dictionary between non-abelian
orbifold data and generalized fluxes has not been worked out.
The non-Abelian nature of the orbifold could possibly allow gauge algebras beyond the
reach of the abelian construction of this paper, including, potentially, gaugings carrying
intrinsic $R$-flux. Extending our dictionary to non-Abelian
orbifold actions is an interesting direction that we leave for future work.	
	
	\section*{Acknowledgments}
    We are gratefull to Ioannis Florakis,
    Nicolas Kovensky, Giorgio Leone, Diego Marqués, Miguel Montero, Hector Parra de Freitas, Alessandro Tomasiello and Daniel Waldram for useful comments and insightful discussions. The authors are particularly thankful to Carlo Angelantonj and Giuseppe Dibitetto for their comments and suggestions.
    D.P. would like to thank the Physics Department in Torino for hospitality during various stages of this work.
	\appendix
	\section{Conventions, fluxes and modular functions}
\label{App:conventions}

\subsection{Conventions}
We use the following notation for the indices:
\begin{equation}
\label{indices}
    \begin{split}
        &M,N,...  \text{ 10D spacetime indices}
        \\
        &\mu,\nu,... \ \ \ \text{7D spacetime indices}
        \\
        &m,n,... \ \text{ SL(4) indices}
        \\
        &\bar{m},\bar{n},... \ \text{ SO(4) indices}
        \\
        &A,B,... \ \text{ SO(3,3) indices}
        \\
        &a,b,... \ \text{ SO(3) indices}
        \\
        &i,j,... \ \ \ \text{ SU(2)$_R$ indices}
        \\
        &\hat{i},\hat{j},...\ \ \ \text{ SU(2) indices}
        \\
        &I,J,... \ \ \text{ T$^d$ indices }
    \end{split}
\end{equation}

\noindent
We can now introduce the ’t Hooft symbols $(G_A)^{mn}$ \cite{dibitetto_all_2015}. These objects are invariant tensors that maps the
fundamental representation of $\mathrm{SO}(3,3)$ 
to the antisymmetric two-form representation of $\mathrm{SL}(4)$ as:
\begin{equation}
v^{mn} = (G_A)^{mn}\, v^A ,
\end{equation}
where $v^A$ transforms as a vector of $\mathrm{SO}(3,3)$ and
$v^{mn} = - v^{nm}$.
The two-form representation of $\mathrm{SL}(4)$ is real due to the
existence of the Levi-Civita tensor, which relates $v^{mn}$ to its dual:
\begin{equation}
v_{mn} = \frac{1}{2} \,\varepsilon_{mnpq}\, v^{pq}.
\end{equation}
The inverse map is then implemented by the "inverse" 't Hooft symbol:
\begin{equation}
(G_A)_{mn} \equiv \frac{1}{2} \varepsilon_{mnpq} (G_A)^{pq}.
\end{equation}
Given the light-cone $\mathrm{SO}(3,3)$ metric $\eta_{AB}$ and the six-dimensional Levi--Civita symbol $\varepsilon_{ABCDEF}$, the 't Hooft symbols $(G_A)_{mn}$ and $(G_A)^{mn}$ satisfy the identities:
\begin{align}
(G_A)_{mn} (G_B)^{nm} &= 2 \eta_{AB}, \\
(G_A)_{mp} (G_B)^{pn} + (G_B)_{mp} (G_A)^{pn}
&= - \delta_m^{\, n} \, \eta_{AB}, \\
(G_{\left[A\right.}  )_{mp} (G_B)^{pq} (G_C)_{qr}
(G_D)^{rs} (G_E)_{st} (G_{\left.F\,\right]})^{tn}
&= -\frac{1}{8}\delta_m^{\, n} \, \varepsilon_{ABCDEF}.
\end{align}
For concreteness we adopt the following explicit representation for the ’t Hooft symbols
in light-cone coordinates:

\begin{equation}
\begin{split}
(G_1)^{mn} &=
\begin{pmatrix}
0 & -1 & 0 & 0 \\
1 &  0 & 0 & 0 \\
0 &  0 & 0 & 0 \\
0 &  0 & 0 & 0
\end{pmatrix}, &
(G_2)^{mn} &=
\begin{pmatrix}
0 & 0 & -1 & 0 \\
0 & 0 &  0 & 0 \\
1 & 0 &  0 & 0 \\
0 & 0 &  0 & 0
\end{pmatrix}, &
\\[1em]
(G_3)^{mn} &=
\begin{pmatrix}
0 & 0 & 0 & -1 \\
0 & 0 & 0 &  0 \\
0 & 0 & 0 &  0 \\
1 & 0 & 0 &  0
\end{pmatrix},&
(G_{\bar1})^{mn} &=
\begin{pmatrix}
0 & 0 & 0 & 0 \\
0 & 0 & 0 & 0 \\
0 & 0 & 0 & -1 \\
0 & 0 & 1 & 0
\end{pmatrix}, &
\\[1em]
(G_{\bar2})^{mn} &=
\begin{pmatrix}
0 & 0 & 0 & 0 \\
0 & 0 & 0 & 1 \\
0 & 0 & 0 & 0 \\
0 & -1 & 0 & 0
\end{pmatrix}, &
(G_{\bar3})^{mn} &=
\begin{pmatrix}
0 & 0 & 0 & 0 \\
0 & 0 & -1 & 0 \\
0 & 1 & 0 & 0 \\
0 & 0 & 0 & 0
\end{pmatrix}.   
\end{split}
\end{equation}

\noindent
Let us also fix our conventions for what concerns the $SO(4)$ Dirac matrices.
We adopt a Weyl representation for the Dirac matrices,
so they take the block form
\begin{equation}
\mathbf{\Gamma}_{\bar{m}} =
\begin{pmatrix}
0_2 & (\Gamma_{\bar{m}})^{i \hat{i}} \\
(\bar\Gamma_{\bar{m}})_{\hat{i} i} & 0_2
\end{pmatrix},
\end{equation}
which satisfies the Clifford algebra:
\begin{equation}
\{\mathbf{\Gamma}_{\bar{m}} , \mathbf{\Gamma}_{\bar{n}}\}
= 2 \delta_{\bar{m}\bar{n}} \mathds{1}_4 .
\end{equation}
Explicitly the gamma blocks given by:
\begin{equation}
    \begin{split}
        \Gamma_{\bar{m}}&=\left(   \mathds{1}_2,i \sigma^1,i \sigma^3,i \sigma^3 \right)_{\bar{m}}
        \\
        \bar{\Gamma}_{\bar{m}}&=\left(   \mathds{1}_2,-i \sigma^1,-i \sigma^3,-i \sigma^3 \right)_{\bar{m}}
    \end{split}
\end{equation}
Here $\{\sigma^i\}_{i=1,2,3}$ are the Pauli matrices
\begin{align}
\sigma^1 &=
\begin{pmatrix}
0 & 1 \\
1 & 0
\end{pmatrix}, &
\sigma^2 &=
\begin{pmatrix}
0 & -i \\
i & 0
\end{pmatrix}, &
\sigma^3 &=
\begin{pmatrix}
1 & 0 \\
0 & -1
\end{pmatrix}.
\end{align}

\subsection{Scherk-Schwarz reduction}
\label{app:fluxes}
In this subsection we put together our parametrisation of the twist matrix and the fluxes in a generalized Scherk-Schwarz reduction.
(Over) parametrizing the twist matrix as:
\begin{equation}
     U^A{}_M = \left(\begin{matrix} u_a{}^m & u_a{}^n b_{nm} \\ u^a{}_n
\beta^{nm} & u^a{}_m + u^a{}_n \beta^{np}
b_{pm}\end{matrix}\right)\, ,\label{TwistParam}
\end{equation}
\noindent
the fluxes \eqref{commutators} have the following explicit form \cite{aldazabal_double_2013}:
\begin{equation}
    \begin{split}
        H_{abc} &= 3\left[ \nabla_{[a} b_{bc]}  - b_{d[a} \tilde\nabla^d b_{bc]}\right]\, ,
        \\
        \omega_{ab}{}^{c} &= 2 \Gamma_{[ab]}{}^c + \tilde \nabla^c b_{ab} + 2
        \Gamma^{mc}{}_{[a} b_{b]m} + \beta^{cm}H_{mab}\, ,
        \\
        Q_c{}^{ab} &= 2 \Gamma^{[ab]}{}_c +\partial_c\beta^{ab}+  b_{cm} \tilde
        \partial^m\beta^{ab} + 2 \omega_{mc}{}^{[a}\beta^{b]m}  -H_{mnc}
        \beta^{ma}\beta^{nb}\, ,
        \\
        R^{abc} &= 3\left[\beta^{[\underline am}\nabla_m \beta^{\underline b\underline
        c]}
        + \tilde \nabla^{[a} \beta^{bc]} +
        b_{mn} \tilde \nabla^n\beta^{[ab} \beta^{c]m} + \beta^{[\underline
        am}\beta^{\underline bn} \tilde\nabla^{\underline
        c]}b_{mn}
        \right ] +
        \beta^{am}\beta^{ bn}\beta^{cl}H_{mnl}\, ,
    \end{split}
\label{fluxes}
\end{equation}
\noindent
with:
\begin{equation}
    \begin{split}
        &u_a{}^m u^a{}_n = \delta^m_n \ , \ \ \ \ \ \ u_a{}^m u^b{}_m = \delta_a^b\,
        ,\quad
        b_{ab} = u_a{}^m u_b{}^n b_{mn} \ , \ \ \ \ \ \ \beta^{ab} = u^a{}_m u^b{}_n
        \beta^{mn}\, ,
        \\
        &\partial_a = u_a{}^m \partial_m  \ , \ \ \ \ \ \ \tilde \partial^a = u^a{}_m
        \tilde \partial^m\, ,
        \\
        &\nabla_a b_{bc}=\partial_a b_{bc}-\Gamma_{ab}{}^d b_{dc}-\Gamma_{ac}{}^d
        b_{bd}\, ,\quad
        \tilde \nabla^a b_{bc}=\tilde\partial^a b_{bc}+\Gamma^{ad}{}_{b}
        b_{dc}+\Gamma^{ad}{}_{c} b_{bd}\, ,
        \\
        &\nabla_a
        \beta^{bc}=\partial_a\beta^{bc}+\Gamma_{ad}{}^b\beta^{dc}+\Gamma_{ad}{}^c\beta^{
        bd}\, ,\quad
        \tilde \nabla^a
        \beta^{bc}=\tilde\partial^a\beta^{bc}-\Gamma^{ab}{}_{d}\beta^{dc}-\Gamma^{ac}{}_
        {d}\beta^{bd}\, ,
        \\
        &\Gamma_{ab}{}^c = u_a{}^m \partial_m u_b{}^n u  ^c{}_n \ , \ \ \ \ \ \
        \\
        &\Gamma^{ab}{}_c = u^a{}_m \tilde \partial^m u^b{}_n u_c{}^n\, .\label{Gammas}
    \end{split}
\end{equation}

\subsection{Modular functions}
Let us first introduce the Dedekind eta and Jacobi theta functions and their modular properties, see for example \cite{angelantonj_open_2002}.
The definitions are the following:
\begin{align}
    &\eta(\tau)=q^\frac{1}{24} \prod_{n=1}^{\infty} (1-q^n),
    \\
    &\vartheta\left[\genfrac{}{}{0pt}{}{\alpha}{\beta}\right]
    (\tau)
    =
    e^{2\pi i \alpha \beta} \,
    q^{\alpha^2/2}
    \prod_{n=1}^{\infty}
    (1 - q^n)
    \prod_{n=1}^{\infty}
    \left(1 + q^{n+\alpha-\frac{1}{2}} e^{2\pi i \beta}\right)
    \left(1 + q^{n-\alpha-\frac{1}{2}} e^{-2\pi i \beta}\right).
\end{align}
Under the modular transformations $S:\tau\rightarrow-1/\tau$ and $T:\tau\rightarrow\tau+1$ the above functions transforms as:
\begin{align}
        &\eta(-1/\tau)=\sqrt{-i\tau}\ \eta(\tau),
        \\
        &\eta\left(\tau+1\right)=e^{-i\frac{\pi}{12}}\ \eta(\tau),
        \\&\vartheta\left[\genfrac{}{}{0pt}{}{\alpha}{\beta}\right]\!(-1/\tau)
        = \sqrt{-i\tau}\,e^{2\pi i\,\alpha\beta}\,
        \vartheta\left[\genfrac{}{}{0pt}{}{\beta}{-\alpha}\right]( \tau),
        \\
       &\vartheta\left[\genfrac{}{}{0pt}{}{\alpha}{\beta}\right] (\tau+1) = e^{-i\pi\,\alpha(\alpha-1)}\,
    \vartheta\left[\genfrac{}{}{0pt}{}{\alpha}{ \beta+\alpha-\nicefrac12}\right](\tau)
    .
\end{align}
Let us also define the four Jacobi constants: 
$\vartheta_1=\vartheta\left[\genfrac{}{}{0pt}{}{\nicefrac{1}{2}}{\nicefrac{1}{2}}\right]$,
$\vartheta_2=\vartheta\left[\genfrac{}{}{0pt}{}{\nicefrac{1}{2}}{0}\right]$,
$\vartheta_3=\vartheta\left[\genfrac{}{}{0pt}{}{0}{0}\right]$,
$\vartheta_4=\vartheta\left[\genfrac{}{}{0pt}{}{0}{\nicefrac{1}{2}}\right]$,
and also two useful relations:
\begin{equation}
    \vartheta_2\vartheta_3\vartheta_4=2\eta^3, 
    \qquad
    \theta_3^4-\theta_4^4-\theta_2^4=0.
\end{equation}
We also introduce the $SO(2n)$ affine characters:
\begin{equation}
   \begin{split}
    O_{2n}&=\frac{1}{2\eta^{n}}\left(\vartheta_3^{n}+\vartheta_4^{n}\right), 
    \\
    V_{2n}&=\frac{1}{2\eta^{n}}\left(\vartheta_3^{n}-\vartheta_4^{n}\right), 
    \\
    S_{2n}&=\frac{1}{2\eta^{\,n}}\left(\vartheta_2^{n}+i^{-n}\vartheta_1^{\,n}\right), 
    \\
    C_{2n}&=\frac{1}{2\eta^{\,n}}\left(\vartheta_2^{\,n}-i^{-n}\vartheta_1^{\,n}\right).
    \end{split}
\end{equation}

\subsection{Correlation functions}

In this section we introduce the basic correlation functions on the sphere which we use in the evaluation of the tree level amplitudes.
In our conventions, which are the same of \cite{Blumenhagen:2013fgp} with $\alpha'=1$, the correlation function of two spacetime bosons is
\begin{equation}
    \braket{X^\mu(z,\bar z) X^\nu(\omega,\bar \omega)}=-\frac{1}{2}\eta^{\mu \nu}\log|z-\omega|^2,
\end{equation}
out of which one can obtain the correlation functions:
\begin{equation}
    \braket{\partial X^\mu(z)X^\nu(\omega)}=-\frac{1}{2}\frac{\eta^{\mu \nu}}{(z-\omega)}, 
    \qquad\braket{\partial X^\mu(z)\partial X^\nu(\omega)}=\frac{1}{2}\frac{\eta^{\mu \nu}}{(z-\omega)^2},
\end{equation}
and analogous expressions for the anti-holomorphic derivatives. Using the $X^\mu$'s one can also construct the conformal field $e^{ik\cdot X}(z,\bar z)$, where the normal ordering is understood. The correlation function for $n$ such fields is:
\begin{equation}
    \braket{\prod_{i=1}^n\ e^{ik_{i}\cdot X}(z_i,\bar z_i)}=\prod_{j<l} \left|z_j-z_l\right|^{ k_j\cdot k_l} \ \delta(\Sigma_{i}\,   k^\mu_i).
\end{equation}
For our purposes we also need $\braket{\partial X^\mu(0) e^{ik_1\cdot X}(0,0)e^{ik_2\cdot X}(z_\infty,\bar{z}_\infty)e^{ik_3\cdot X}(1,1)}$, which is
\begin{equation}
    \begin{split}
        &=\frac{i}{2}\left(\frac{k^\mu_2}{z_\infty}+k_3^\mu\right)\braket{\prod_{i=1}^n\ e^{ik_{i}\cdot X}(z_i,\bar z_i)}\\
        &=\frac{i}{2}\left(\frac{k^\mu_2}{z_\infty}+k_3^\mu\right) |0-z_\infty|^{k_1\cdot k_3}|0-1|^{k_1\cdot k_2}|z_\infty-1|^{k_2\cdot k_3}\ \delta(k)\\
        &\simeq\frac{i}{2}\left(\frac{k^\mu_2}{z_\infty}+k_3^\mu\right) |z_\infty|^{k_2\cdot (k_1+k_3)}\ \delta(k)\\
        &=\frac{i}{2}\left(\frac{k^\mu_2}{z_\infty}+k_3^\mu\right) |z_\infty|^{-k_2^2 }\ \delta(k).
    \end{split}
    \label{derE}
\end{equation}
Notice that $\simeq$ symbol means that the equality holds in the regime $z_\infty \gg 1$; we also used the  momentum conservation. All the above expressions can be applied to a set of compact bosons by just using the euclidean metric and allowing for left momenta different form the right momenta in general. Let us just write down an example of the analogous of \eqref{derE} for compact bosons with different left and right moving momenta in the direction 1 (i.e. $y$). For example $\braket{\partial x^a(0) e^{iq_{_L}^a y_{_L}}(0)e^{iq_{_L}^a y_{_L}}(z_\infty)e^{iq_{_L}^a y_{_L}}(1)e^{ip_{_R}^a y_{_R}}(0)e^{ip_{_R}^a y_{_R}}(\bar{z}_\infty)e^{ip_{_R}^a y_{_R}}(1)} $ reads
\begin{equation}
=\frac{i}{2} \, \delta^{a1}\left(\frac{q^b_{_L}}{z_\infty}+q_{_L}^c\right) |z_\infty|^{-(q_{_L}^b)^2 }\delta(q_{_L})\ \delta(q_{_R}) \;
\end{equation}
where we used the conservation deltas, and assumed that the states are level matched.

The fermionic correlation function is
\begin{equation}
    \braket{\psi^\mu(z)\psi^\nu(\omega)}=\frac{\eta^{\mu \nu}}{z-\omega},
\end{equation}
the other correlation function we need can be obtained from this basic correlation trough Wick theorem
\begin{equation}
    \braket{:\psi^\mu\psi^\nu:(0)\, \psi^\rho(z_\infty) \, \psi^{\sigma}(1)}=\left(\frac{\eta^{\nu \rho}\eta^{\mu \sigma}-\eta^{\mu \rho}\eta^{\nu \sigma}}{z_{\infty}}\right),
\end{equation}
where we have made explicit the normal ordering. 
Similar results holds for the anti-holomorphic fermions and in the case of compact fermions. Recall that in our calculations we have a pair of complex fermions in the directions 2 and 3, while in the direction 1 the fermion is real, so the metric for the compact fermions correlator is
\begin{equation}
    S=\begin{pmatrix}
        1 & 0 & 0\\
        0 & 0 & 1\\
        0 & 1 & 0
    \end{pmatrix}.         
\end{equation}

\section{Gauging algebras}
\label{app:algebras}

In this appendix we list the algebras of the groups appearing as gauging in Table \ref{tab:gaugings}.

\subsection*{SO(p,q):}
This class of algebras is one of the most familiar, and consist of special orthogonal groups in signature $\eta_{i j}=\text{diag}(+1_{p},-1_{q})$, and consists of $\frac{(p+q)(p+q-1)}{2}$ elements $M_{ij}=-M_{ji}$, with $i,j,=1,...,p+q$,  satisfying:
\begin{equation}
    [M_{ij}, M_{kl}]= 
    \eta_{jk} M_{il}
    - \eta_{ik} M_{jl}
    - \eta_{jl} M_{ik}
    + \eta_{il} M_{jk}. \\
\end{equation}

\subsection*{ISO(p,q):}
This class is the affine extension of $SO(p,q)$ by translations $P_i$, with commutation relations:
\begin{equation}
    \begin{split}
    [M_{ij}, M_{kl}]&= 
    \eta_{jk} M_{il}
    - \eta_{ik} M_{jl}
    - \eta_{jl} M_{ik}
    + \eta_{il} M_{jk}. \\
    [M_{ij}, P_k] &= 
    \eta_{jk} P_i
    - \eta_{ik} P_j.
    \end{split}
\end{equation}
The group has dimension  $\frac{(p+q)(p+q+1)}{2}$.

\subsection*{CSO(p,q,r):}
In this case we have a contraction of $SO(p+r,q)$ in a way that the $p+r+q$ dimensional metric becomes singular along r diretions. In this case is better to split the generators  as follows: $M_{i j}=-M_{j i}$, $P_{i,m}$, $Z_{mn}=-Z_{nm}$, with $m,n=p+q+1,...,p+q+r$. The commutation relations reads:
\begin{equation}
    \begin{split}
    [M_{ij}, M_{kl}]&= 
    \eta_{jk} M_{il}
    - \eta_{ik} M_{jl}
    - \eta_{jl} M_{ik}
    + \eta_{il} M_{jk}.
    \\[6pt]
    [M_{ij}, P_{k,m}] &= 
    \eta_{jk} P_{i,m}
    - \eta_{ik} P_{j,m}.
    \\
    [P_{i,m}, P_{j,n}] &= \eta_{i j}Z_{mn},
    \end{split}
\end{equation}

\noindent
and the group has dimension  $\frac{(p+r+q)(p+r+q-1)}{2}$. Notice that $CSO(p,q,1)=ISO(p,q)$ and $CSO(p,q,0)=SO(p,q)$.

\subsection*{Solvable and Nilpotent algebras:}

Here we describe the solvable and nilpotent algebras appearing as gaugings in Table \ref{tab:gaugings}. 
There are only two nilpotent algebras as gaugings, one is $CSO(1,0,3)$, the other one we call $\mathfrak{n}_6$. Our notation for the solvable algebras that are not  $CSO(p,q,r)$\footnote{$CSO(2,0,2)$ and $CSO(1,1,2)$ are solvable.} is $\mathfrak{s}^{\lambda}_{d,i^{\pm}}$, where $d$ stands for the dimension of the algebra, $i$ is the label we assign to the particular algebra, and $\pm$ refers to signs in the structure constants. Moreover, one algebra is a continuous family depending on one parameter $\lambda$.

All the solvable algebras have a dimension 5 nilradical. 
In \cite{a_shabanskaya_six-dimensional_2013} the authors classify the indecomposable six-dimensional algebras with 5-dimensional nilradical, and also treat some of the 4- and 5-dimensional indecomposable algebras that we present in this appendix. We therefore use their notation in what follows. 
We denote the algebras generators as $\{e_i\}_{i=1,\cdots,6}$.

\paragraph{Nilpotent and decomposable algebras.}
Here we list the only nilpotent and the two decomposable algebras appearing as gaugings.

\subsubsection*{$\mathfrak{n}_6=A_{5,3}\oplus A_1$:}
This is the only nilpotent algebra, it consist of an abelian factor generated by $e_6$ and the five-dimensional indecomposable algebra, denoted $A_{5,3}$ in \cite{a_shabanskaya_six-dimensional_2013}:
\begin{equation}
[e_2,e_4]=e_3,\qquad
[e_2,e_5]=e_1,\qquad
[e_4,e_5]=e_2.
\end{equation}

\subsubsection*{$\mathfrak{s}_{6,2^{\pm}}$:}
These are two decomposable solvable algebras with two abelian factors generated by $\{e_5,e_6\}$ and the four-dimensional indecomposable algebras:
\begin{equation}
\begin{aligned} 
\relax 
[e_2,e_3]&=e_1,
&
[e_2,e_4]&=-e_3,
&
[e_3,e_4]&=\pm e_2.
\end{aligned}
\end{equation}
depending on the sign of the last commutation relation.

\paragraph{Algebras with \texorpdfstring{$A_{5,4}$}{} nilradical.}

All the following algebras have as nilraddical the five-dimensional Heisemberg algebra $A_{5,4}$, and have the following commutation relations. We identify them in the classification of  \cite{a_shabanskaya_six-dimensional_2013} that uses the labels $\mathfrak g_{6,n}$.
\subsubsection*{$\mathfrak{s}_{6,1^{+}}=\mathfrak g_{6,93}(\delta=0,a=1)$:}
\begin{equation}
\begin{aligned}
\relax 
[e_2,e_4]&=e_1,
&\qquad
[e_3,e_5]&=e_1,
\\
[e_2,e_6]&=e_4+e_5,
&
[e_3,e_6]&= e_4,
\\
[e_4,e_6]&=-e_3,
&
[e_5,e_6]&=-e_3-e_2.
\end{aligned}
\end{equation}

\subsubsection*{$\mathfrak{s}_{6,1^{-}}=\mathfrak g_{6,83}(\delta=0,a=1)$:}
\begin{equation}
\begin{aligned}\relax 
[e_2,e_4]&=e_1,
&\qquad
[e_3,e_5]&=e_1,
\\
[e_2,e_6]&=e_2+e_3,
&
[e_3,e_6]&=e_3,
\\
[e_4,e_6]&=-e_4,
&
[e_5,e_6]&=-e_4-e_5.
\end{aligned}
\end{equation}

\subsubsection*{$\mathfrak{s}_{6,3^{+}}^{b}=\mathfrak{g}'_{6,92}(\delta=0,a=1,b=\frac{|\alpha|-|\beta|}{|\alpha|+|\beta|})$:}

\begin{equation}
\begin{aligned}\relax 
[e_2,e_4]&=e_1,
&
[e_3,e_5]&=e_1,
\\
[e_2,e_6]&=e_4,
&
[e_3,e_6]&=b e_5,
\\
[e_4,e_6]&=-e_2,
&
[e_5,e_6]&=-be_3.
\end{aligned} 
\end{equation}

\subsubsection*{$\mathfrak{s}_{6,3^{-}}^{a}=\mathfrak{g}_{6,82}(\delta=0,a=\frac{|\alpha|-|\beta|}{|\alpha|+|\beta|},b=1)$:}

\begin{equation}
\begin{aligned}\relax 
[e_2,e_4]&=e_1,
&
[e_3,e_5]&=e_1,
\\
[e_2,e_6]&=ae_2,
&
[e_3,e_6]&=\,e_3,
\\
[e_4,e_6]&=-a e_4,
&
[e_5,e_6]&=-e_5.
\end{aligned} 
\end{equation}

\subsubsection*{$\mathfrak{s}_{6,4}=\mathfrak{g}_{6,88}(\alpha=0,a=1)$:}
\begin{equation}
\begin{aligned}\relax 
[e_2,e_4]&=e_1,
&
[e_3,e_5]&=e_1,
\\
[e_6,e_2]&=-e_2-e_3,
&
[e_6,e_3]&=e_2-e_3,
\\
[e_6,e_4]&=e_4-e_5,
&
[e_6,e_5]&=e_4+e_5.
\end{aligned}
\end{equation}

\paragraph{Focus on the algebras appearing in 6:}
For the sake of clarity we rewrite the algebra appearing in the gauging 6 in table \ref{tab:gaugings} ($\mathfrak{g}'_{6,92}$) in terms of the $Z$ and $\chi$ generators:
\begin{equation}
    \begin{aligned}
    &[Z_1,Z_2]=\alpha \chi^3 + \beta Z_3,
    \qquad
    [Z_1,Z_3]=-\alpha \chi^2 -\beta Z_2,
    \\
    &
    [Z_1,\chi^2]=\beta \chi^3 + \alpha Z_3,
    \qquad
    [Z_1,\chi^3]=-\beta \chi^2 - \alpha Z_2,
    \\
    &
    [Z_2,Z_3]= \alpha \chi^1,
    \qquad
    [Z_2,\chi^3]=\beta \chi^1,
    \qquad
    [Z_3,\chi^2]=-\beta \chi^1,
    \qquad
    [\chi^2,\chi^3]=\alpha \chi^1.
    \end{aligned}
\end{equation} 

\section{Symmetries of the Narain Lattice}

\label{App:Narain_Sym}

In this appendix we elaborate on the symmetries of the Narain Lattice and fix our conventions. 
Let us consider the Narain Lattice $\Gamma_{d,d}(E)$ in $d$-dimensions with moduli $E$. The left and right momenta of our lattice are
\begin{equation}
    \begin{cases}
        \left(P_L\right)_I=\left(e^{-1}\right)_{I}^{\ i}\left( m_i +\left(G_{ij}-B_{ij}\right)n^{j}  \right)=\left(e^{-1}\right)_{I}^{\ i}\left( m_i +E^t_{ij} n^{j}  \right),
        \\
        \\
        \left(P_R\right)_I=\left(e^{-1}\right)_{I}^{\ i}\left( m_i -\left(G_{ij}+B_{ij}\right)n^{j}  \right)=\left(e^{-1}\right)_{I}^{\ i}\left( m_i -E_{ij} n^{j}  \right),
    \end{cases}
\end{equation}
where $m_i,n^j\in \mathbb{Z}$ are the momenta and winding on the torus, $e$ is the lattice vielbein, such that $G_{ij}=e_i^{\ I} \delta_{IJ}  e^J_{\ j}=\vec{e_i}\cdot \vec{e_j}^t$, and $E_{ij}=G_{ij}+B_{ij}$.
For the sake of simplicity from now on we employ a matrix notation as follows:
\begin{equation}
    P=\left(\genfrac{}{}{0pt}{}{P_L}{P_R}\right)=
    \begin{pmatrix}
        e^{-1}& 0\\
        0&e^{-1}
    \end{pmatrix}
    \begin{pmatrix}
        \mathds{1}& E^t\\
        \mathds{1}& -E
    \end{pmatrix}
    \left(\genfrac{}{}{0pt}{}{\vec{m}}{\vec{n}}\right) \in \Gamma_{d,d}(E)
\end{equation}
\noindent
For a matrix $\theta$ acting on $P$ to be a symmetry of the Narain Lattice we need $P'=\theta \ P\in \Gamma_{d,d}(E)$. This translates in the following general form for $P'$:
\begin{equation}
    P'=
    \begin{pmatrix}
        e^{-1}& 0\\
        0&e^{-1}
    \end{pmatrix}
    \begin{pmatrix}
        \mathds{1}& E^t\\
        \mathds{1}& -E
    \end{pmatrix}
    \begin{pmatrix}
        a & b\\
        c & d
    \end{pmatrix}
    \left(\genfrac{}{}{0pt}{}{\vec{m}}{\vec{n}}\right),
    \  \ \mathcal{O}=\begin{pmatrix}
        a & b\\
        c & d
    \end{pmatrix} \in O(d,d;\mathbb{Z}),
\end{equation}
which means that $\theta$ preserves the moduli and acts just as an integral transformation on the momenta and winding preserving the signature of the lattice. The constraints on the $d\times d$ integral blocks of $\mathcal{O}$ coming form the signature preserving condition $\mathcal{O}^t\eta\mathcal{O}=\eta$ are:
\begin{equation}
        a^tc+c^ta=0,
        \qquad
        d^tb+b^td=0,
        \qquad
        a^td+c^tb=\mathds{1}.
\label{eq:oddcond}
\end{equation}
Let's reconstruct $\theta$ from $P'$: 
\begin{equation}
    \begin{split}
        P'&=
            \begin{pmatrix}
                e^{-1}& 0\\
                0&e^{-1}
            \end{pmatrix}
            \begin{pmatrix}
                \mathds{1}& E^t\\
                \mathds{1}& -E
            \end{pmatrix}
            \begin{pmatrix}
                a & b\\
                c & d
            \end{pmatrix}
            \left(\genfrac{}{}{0pt}{}{\vec{m}}{\vec{n}}\right)
        \\ \\
        &=\begin{pmatrix}
        e^{-1}& 0\\
        0&e^{-1}
        \end{pmatrix}
        \begin{pmatrix}
            \mathds{1}& E^t\\
            \mathds{1}& -E
        \end{pmatrix}
        \begin{pmatrix}
            a & b\\
            c & d
        \end{pmatrix}
        \begin{pmatrix}
            \mathds{1}& E^t\\
            \mathds{1}& -E
        \end{pmatrix}^{-1}
        \begin{pmatrix}
            e& 0\\
            0&e
        \end{pmatrix}
        \begin{pmatrix}
            e^{-1}& 0\\
            0&e^{-1}
        \end{pmatrix}
        \begin{pmatrix}
            \mathds{1}& E^t\\
            \mathds{1}& -E
        \end{pmatrix}
        \left(\genfrac{}{}{0pt}{}{\vec{m}}{\vec{n}}\right)
        \\ \\
        &=\begin{pmatrix}
        e^{-1}& 0\\
        0&e^{-1}
        \end{pmatrix}
        \begin{pmatrix}
            \mathds{1}& E^t\\
            \mathds{1}& -E
        \end{pmatrix}
        \begin{pmatrix}
            a & b\\
            c & d
        \end{pmatrix}
        \frac{1}{2}
        \begin{pmatrix}
            E& E^t\\
            \mathds{1}& -\mathds{1}
        \end{pmatrix}
        \begin{pmatrix}
            G^{-1}& 0\\
            0&G^{-1}
        \end{pmatrix}
        \begin{pmatrix}
            e& 0\\
            0&e
        \end{pmatrix}
        P
        \\ \\
        &\equiv 
        \begin{pmatrix}
            e^{-1}& 0\\
            0&e^{-1}
        \end{pmatrix}
        M
        \begin{pmatrix}
            e& 0\\
            0&e
        \end{pmatrix}
        P
        \\ \\
        &\equiv 
        \theta
        P
    \end{split}
\end{equation}
In the last step we defined for simplicity an intermediate matrix $M$ conjugate to $\theta$. Let us expand $M$:
\begin{equation}
    \begin{split}
        M&=\frac{1}{2}\begin{pmatrix}
            \mathds{1}& E^t\\
            \mathds{1}& -E
        \end{pmatrix}
        \begin{pmatrix}
            a & b\\
            c & d
        \end{pmatrix}
        \begin{pmatrix}
            E& E^t\\
            \mathds{1}& -\mathds{1}
        \end{pmatrix}
        \begin{pmatrix}
            G^{-1}& 0\\
            0&G^{-1}
        \end{pmatrix}
     \\ \\
        &=\frac{1}{2}
        \begin{pmatrix}
            (aE+b)+E^t(cE+d)& (aE^t-b)+E^t(cE^t-d)\\
            (aE+b)-E(cE+d)& (aE^t-b)-E(cE^t-d)
        \end{pmatrix}
        \begin{pmatrix}
            G^{-1}& 0\\
            0&G^{-1}
        \end{pmatrix}.   
    \end{split}
    \label{eq:M}
\end{equation}
\noindent
Since we are interested in a rotation $\theta$ that is a symmetry of the full CFT, in view of quotienting it, we impose that it must not mix the left and right sectors . This reflects in the same condition on $M$ since $\theta$ and $M$ are related by diagonal matrices, so we impose:
\begin{equation}
    M\overset{!}{=}
    \begin{pmatrix}
        M_L&0\\
        0&M_R
    \end{pmatrix}.
\end{equation}
Imposing this to \eqref{eq:M} one obtains:
\begin{equation}
    E =(aE+b)(cE+d)^{-1},
    \qquad 
    E^t=-(aE^t-b)(cE^t-d)^{-1};
    \label{eq:Econd}
\end{equation}
which are the usual relations for the moduli,  and it is easy to show that are equivalent using the defining properties of $O(d,d;\mathbb{Z})$ in \eqref{eq:oddcond}.
Using \eqref{eq:Econd} one can rewrite $M$ as:
\begin{equation}
    M=\begin{pmatrix}
        G(d+cE)G^{-1}&0\\
        0&G(d-cE^t)G^{-1}
    \end{pmatrix},
\end{equation}
and finally the expression for $\theta$ in terms of the moduli and the sub-matrices of $\mathcal{O}$ is:
\begin{equation}
    \theta=\begin{pmatrix}
        \theta_L&0\\
        0&\theta_R
    \end{pmatrix}
    =
    \begin{pmatrix}
        e^t(d+cE)e^{-t}&0\\
        0&e^t(d-cE^t)e^{-t}
    \end{pmatrix}.
\end{equation}
This means that $\theta$ is similar to  an $O(d,d;\mathbb{Z})$ matrix, which can also be seen from its definition:
\begin{equation}
    \begin{split}
    \theta&=\begin{pmatrix}
            e^{-1}& 0\\
            0&e^{-1}
        \end{pmatrix}
        \begin{pmatrix}
            M_L& 0\\
            0&M_R
        \end{pmatrix}
        \begin{pmatrix}
            e& 0\\
            0&e
        \end{pmatrix}
        \\
        &=
        \begin{pmatrix}
            e^{-1}& 0\\
            0&e^{-1}
        \end{pmatrix}
        \begin{pmatrix}
            \mathds{1}& E^t\\
            \mathds{1}& -E
        \end{pmatrix}
        \begin{pmatrix}
            a& b\\
            c&d
        \end{pmatrix}
        \begin{pmatrix}
            \mathds{1}& E^t\\
            \mathds{1}& -E
        \end{pmatrix}^{-1}
        \begin{pmatrix}
            e& 0\\
            0&e
        \end{pmatrix}
        \\
        &\equiv
        S(E)\mathcal{O}S^{-1}(E),
    \end{split}
\end{equation}
Notice that the similarity condition with an $O(d,d;\mathbb{Z})$ matrix implies in particular that the trace of $\theta$ is integer and that thr eigenvalues of $\theta$ appear in pairs $(\lambda,\lambda^{-1})$:
\begin{equation}
        Tr(\theta)\in\mathbb{Z},        
        \qquad
        \text{Spec}\left(\theta\right)=\left\{ \left(\lambda,\lambda^{-1} \right) | \theta v=\lambda v   \right\}.
\end{equation}
If one defines $\theta_{\pm}=\frac{1}{2}\left(\theta_L\pm\theta_R\right)$ is straightforward to rewrite a,b,c and d in terms of the component of $\theta$:
\begin{equation}
    \begin{cases}
        c\overset{!}{=}e^{-t}\theta_-e^{-1},
        \\
        d\overset{!}{=}e^{-t}\theta_+e^{t}-cB,
        \\
        a\overset{!}{=}e\theta_+e^{-1}+Bc,
        \\
        b\overset{!}{=}e\theta_{-}e^{t}+Bd-aB+BcB.
    \end{cases}
    \label{eq:cond}
\end{equation}
Let us stress that the constraint that $a,b,c$ and $d$ have to be integer matrices is very restrictive and indeed there are just few allowed $\theta$'s, and possible moduli related to a specific choice of $\theta$.
For the specific case of $T^2$ abelian rotations we present the solutions in Table \ref{tab:rotsym}.

\section{Further examples of asymmetric orbifolds}
\label{App:other}

In this appendix we present the $\mathbb{Z}_3$ and $\mathbb{Z}_6$ variant of the $\mathbb{Z}_4$ orbifolds realizing the cases $\beta=\pm \alpha$ in gauging 6. We decided to defere to an appendix these models since they do not add anything but complications to the $\mathbb{Z}_4$ case.

\subsubsection*{$\beta=\alpha$:}

Let us treat the $\mathbb{Z}_3$ and $\mathbb{Z}_6$ cases togheter until it is possible. 
The action on the worldsheet fermions is straightforward, while the one one the bosonic degrees of freedom is a bit more tricky. First of all let us notice that the only 2 dimensional Narain lattice compatible with the orbifold action $(\lambda_L,\lambda_R)=(\frac{k}{3},0)$, with $k=1,2$, is the $SU(3)\times \overline{SU(3)}$ lattice. A possible way to realize this lattice, in our conventions, is to set the lattice metric to be 1/2 the Cartan matrix of $SU(3)$ and the lattice B-field to be 1/2 the antisymmetrized adjacency matrix of $SU(3)$:
\begin{equation}
    \hat{G}=\begin{pmatrix}
        1 & -\frac{1}{2}\\
        -\frac{1}{2}&1
    \end{pmatrix},
    \qquad
    \hat{B}=\begin{pmatrix}
        0 & -\frac{1}{2}\\
        \frac{1}{2}&0
    \end{pmatrix}.
\end{equation}
That said let us introduce the $SU(3)$ chiral  algebra conformal characters as:
\begin{equation}
    \Xi_{l}=\frac{1}{\eta^2}\sum_{m,n\in \mathbb{Z}} q^{\left(m+\frac{l}{3},n+\frac{2l}{3}\right)^t\hat{G}\left(m+\frac{l}{3},n+\frac{2l}{3}\right)},
\end{equation}
with $l=0,1,2$ labelling the three conjugacy classes of the $SU(3)$ algebra; the ones relative to the adjoint, fundamental and anti-fundamental representation respectively.
In terms of these characters, and their anti-holomorphic counterpart, the Narain $SU(3) \times \overline{SU(3)}$ lattice partition function  can be easily written as:
\begin{equation}
    \Lambda_{SU(3)}=\sum_{l=0,1,2} \overline{\Xi}_l \ \Xi_l,
\end{equation}
and this will precisely be the contribution from the compact bosons to our partition function in $\left[\genfrac{}{}{0pt}{}{0}{0}\right]$ sector.
We finally have all the ingredients to build our orbifolds. Let us then analyze the compact bosons contribution to the partition function for the two cases at head:

\paragraph*{$\mathbb{Z}_3$:}
The left moving bosonic oscillator contribution is easily obtained just from the twist explicit expression. The invariant lattice under the twist $(1/3,0)$ is given precisely by the right-moving $SU(3)$ root lattice, while the oscillators are untouched; therefore the right-moving contribution is given by the anti-holomorphic character of the adjoint conjugacy class of $SU(3)$. All togheter the left and right-moving contributions gives:
\begin{equation}
    \Gamma^{(\frac{2}{3})}\left[\genfrac{}{}{0pt}{}{0}{g} \right]= \frac{2 \sin(2\pi g/3)}{\theta\left[\genfrac{}{}{0pt}{}{\nicefrac{1}{2}}{ \nicefrac{1}{2}-\nicefrac{2g}{3}}\right]} \ \overline{\Xi}_0, \qquad g=1,2.
\end{equation}

\paragraph*{$\mathbb{Z}_6$:}
This case is just an alomost trivial extension of the previous case. In practice since the fermions forces the order of the orbifold to be 6 instead of 3 there is one projection sector in which the twist trivializes and therefore the invariant lattice is the whole $SU(3)\times \overline{SU(3)}$ lattice. Explicitely one can write:
\begin{equation}
     \Gamma^{(\frac{1}{3})}\left[ \genfrac{}{}{0pt}{}{0}{g}\right]=
     \left\{
     \begin{split}
        &\sum_{l=0,1,2} \overline{\Xi}_l \ \Xi_l, \qquad \quad \ \ g=0 \mod 3,
        \\ \\
        &\frac{2 \sin(\pi g/3)}{\theta\left[\genfrac{}{}{0pt}{}{\nicefrac{1}{2}}{ \nicefrac{1}{2}-\nicefrac{g}{3}}\right]} \ \overline{\Xi}_0, \qquad g\neq0 \mod3. 
     \end{split}
     \right.
\end{equation}
That said the untwisted sector partition function in both cases reads:
\begin{equation}
	\begin{split}
			\mathcal{Z}^{(\frac{k}{3})}\left[\genfrac{}{}{0pt}{}{0}{g}\right]=&  \
		\frac{1}{2} \sum_{\bar{a},\bar{b}}	\eta_{\bar{a}\bar{b}} \ \frac{\bar{\theta}^4\left[\genfrac{}{}{0pt}{}{\nicefrac{\bar{a}}{2}}{\nicefrac{\bar{b}}{2}}\right]}{\bar{\eta}^4}	
		\ \frac{1}{2} \sum_{a,b}	\eta_{ab} \ \frac{\theta^3\left[\genfrac{}{}{0pt}{}{\nicefrac{a}{2}}{\nicefrac{b}{2}}\right]}{\eta^3}	
		\frac{\theta\left[\genfrac{}{}{0pt}{}{\nicefrac{a}{2}}{\nicefrac{b}{2}+\nicefrac{kg}{3}}\right]}{\eta}	
        \\
        & \quad
	    \Gamma^{(\frac{k}{3})}\left[\genfrac{}{}{0pt}{}{0}{g}\right]\ \Lambda\left[\genfrac{}{}{0pt}{}{0}{g}\right].
	\end{split}
	\label{fasym2}
\end{equation}
The twisted sectors are easily obtained using the already introduced modular properties of the theta functions and the modular transformations of the $SU(3)$ characters:
\begin{equation}
    \Xi_{l}\left(-\frac{1}{\tau}\right)=\sum_{r=0,1,2} S_{k,l} \ \Xi_{l}\left(\tau\right), \qquad
    \Xi_{l}\left(\tau+1\right)=\sum_{r=0,1,2} T_{k,l} \ \Xi_{l}\left(\tau\right), 
\end{equation}
with:
\begin{equation}
    S=\frac{1}{\sqrt{3}}\begin{pmatrix}
        1 & 1 & 1 \\
        1 & e^{2i \pi/3}&e^{-2i \pi/3}\\
        1 & e^{-2i \pi/3}&e^{2i \pi/3}
    \end{pmatrix},
    \qquad
    T=e^{-i \pi/6}\begin{pmatrix}
        1 & 0 & 0 \\
        0 & e^{2i \pi/3}&0\\
        0 & 0&e^{2i \pi/3}
    \end{pmatrix}.
\end{equation}
With the previous relations one can explicitly check the modular invariance of the partition function and the twisted secotrs multiplicities integrality guaranteed by the factor $\sqrt{3}$ in the denominator that cancels the analogous contribution of the sin.

The q-expansion up to massless states can be performed as in the previous cases and one finds:
\begin{equation}
	\mathcal{Z}^{(\frac{k}{3})}\approx\frac{1}{\left(\sqrt{\tau_2}\bar{\eta}\eta\right)^5} \frac{1}{\left(\bar{\eta}\eta\right)^4}
	\left[
	|V_6|^2 + 2 \bar{V}_6 O_6 \ q^\frac{1}{2}- \left(\bar{S}_6+\bar{C}_6\right)V_6 \ \bar{q}^\frac{1}{8}
	\right].
\end{equation}
Which is the same as the $\mathbb{Z}_4$ model, again consiting of the gravity multiplet and a 2 vector multiplets formed by a vector three scalars and a fermion.

\subsubsection*{$\beta=-\alpha$:}
This case is, as in the $\mathbb{Z}_4$ case just a variation of the $\beta=\alpha$ orbifold adding to its generator a $(-1)^{F_R}$. The untwisted partition function reads:
\begin{equation}
	\begin{split}
			\mathcal{Z}^{(\frac{k}{3})}\left[\genfrac{}{}{0pt}{}{0}{g}\right]=&  \
		\frac{1}{2} \sum_{\bar{a},\bar{b}}	\eta_{\bar{a}\bar{b}} \ (-1)^{\bar{a}}\ \frac{\bar{\theta}^4\left[\genfrac{}{}{0pt}{}{\nicefrac{\bar{a}}{2}}{\nicefrac{\bar{b}}{2}}\right]}{\bar{\eta}^4}	
		\ \frac{1}{2} \sum_{a,b}	\eta_{ab} \ \frac{\theta^3\left[\genfrac{}{}{0pt}{}{\nicefrac{a}{2}}{\nicefrac{b}{2}}\right]}{\eta^3}	
		\frac{\theta\left[\genfrac{}{}{0pt}{}{\nicefrac{a}{2}}{\nicefrac{b}{2}+\nicefrac{kg}{3}}\right]}{\eta}	
        \\
        & \quad
	    \Gamma^{(\frac{k}{3})}\left[\genfrac{}{}{0pt}{}{0}{g}\right]\ \Lambda\left[\genfrac{}{}{0pt}{}{0}{g}\right].
	\end{split}
	\label{fasym}
\end{equation}
The full partition function, as before, is obtained by acting with the modular transformations on the untwisted projected seectors.
Finally the q-expansion up to massless states yields:
\begin{equation}
	\mathcal{Z}^{(\frac{k}{3})}\approx\frac{1}{\left(\sqrt{\tau_2}\bar{\eta}\eta\right)^5} \frac{1}{\left(\bar{\eta}\eta\right)^4}
	\left[
	|V_6|^2 + 2 \bar{V}_6 O_6 \ q^\frac{1}{2}- \left(\bar{S}_6+\bar{C}_6\right)O_6 \ q^{\frac{1}{2}} \bar{q}^\frac{1}{8}
	\right].
\end{equation}

\noindent
And the massless spectrum is again the same as the analogous $\mathbb{Z}_4$ models.

	\medskip
    \bibliographystyle{JHEP}
    \bibliography{biblio}
 
\end{document}